\documentclass[twocolumn]{aastex61}

\usepackage{graphicx}
\usepackage{amsmath}
\newcommand{\kms}{km\,s$^{-1}$} 
\newcommand{\hst}{\emph{HST}}   
\newcommand{\tm}{\tablenotemark} \newcommand{\tn}{\tablenotetext}
\newcommand{\hi}{\ion{H}{1}}

\begin{document}

\title{Probing the Southern Fermi Bubble in Ultraviolet Absorption using Distant AGNs}

\shorttitle{The Southern Fermi Bubble in UV Absorption}
\shortauthors{Karim et al.}

\author[0000-0002-5652-8870]{Md Tanveer Karim}
\affiliation{Center for Astrophysics, Harvard University, 60 Garden Street, MS 10, Cambridge, MA 02138, USA}
\affiliation{Maria Mitchell Observatory, Maria Mitchell Association, MA 02554, USA}
\affiliation{Department of Physics and Astronomy, University of Rochester, NY 14627, USA}
\author[0000-0003-0724-4115]{Andrew J. Fox}
\affiliation{Space Telescope Science Institute, 3700 San Martin Drive, Baltimore, MD 21218, USA}
\author[0000-0003-1892-4423]{Edward B. Jenkins}
\affiliation{Princeton University Observatory, Princeton, NJ 08544, USA}
\author{Rongmon Bordoloi}
\affiliation{MIT-Kavli Center for Astrophysics and Space Research, 77 Massachusetts Avenue, Cambridge, MA, 02139}
\author[0000-0002-0507-7096]{Bart P. Wakker}
\affiliation{University of Wisconsin--Madison, 475 North Charter St., Madison, WI 53706}
\author{Blair D. Savage}
\affiliation{University of Wisconsin--Madison, 475 North Charter St., Madison, WI 53706}
\author[0000-0002-6050-2008]{Felix J. Lockman}
\affiliation{National Radio Astronomy Observatory, P.O. Box 2, Rt. 28/92, Green Bank, WV 24944}
\author{Steven M. Crawford}
\affiliation{South African Astronomical Observatory, 7935 Cape Town, South Africa}
\author{Regina A. Jorgenson}
\affiliation{Maria Mitchell Observatory, Maria Mitchell Association, MA 02554, USA}
\author[0000-0001-7516-4016]{Joss Bland-Hawthorn}
\affiliation{Sydney Institute for Astronomy, School of Physics, University of Sydney, NSW 2006, Australia}
\correspondingauthor{Md Tanveer Karim}
\email{tanveer.karim@cfa.harvard.edu}

\begin{abstract}
The Fermi Bubbles are two giant gamma-ray emitting lobes extending 55$\degr$ above and below the Galactic Center. While the Northern Bubble has been extensively studied in ultraviolet (UV) absorption, little is known about the gas kinematics of the southern Bubble. We use UV absorption-line spectra from the Cosmic Origins Spectrograph (COS) on the {\it Hubble Space Telescope} to probe the southern Fermi Bubble using a sample of 17 background AGN projected behind or near the Bubble. We 
measure the incidence of high-velocity clouds (HVC), finding that four out of six 
sightlines passing through the Bubble show HVC absorption, versus 
six out of eleven passing outside. 
We find strong evidence that the maximum absolute LSR velocity of the HVC components decreases as a function of galactic latitude within the Bubble, for both blueshifted and redshifted components, as expected for a decelerating outflow. We explore whether
the column-density ratios \ion{Si}{4}/\ion{Si}{3}, \ion{Si}{4}/\ion{Si}{2} and \ion{Si}{3}/\ion{Si}{2} correlate with the absolute galactic latitude within the Bubble. 
These results demonstrate the use of UV absorption-line spectroscopy to characterize the kinematics and ionization conditions of embedded clouds in the Galactic Center outflow.
\end{abstract}


\section{Introduction}
\label{sec:intro}

\begin{figure*}[!ht]
    \plotone{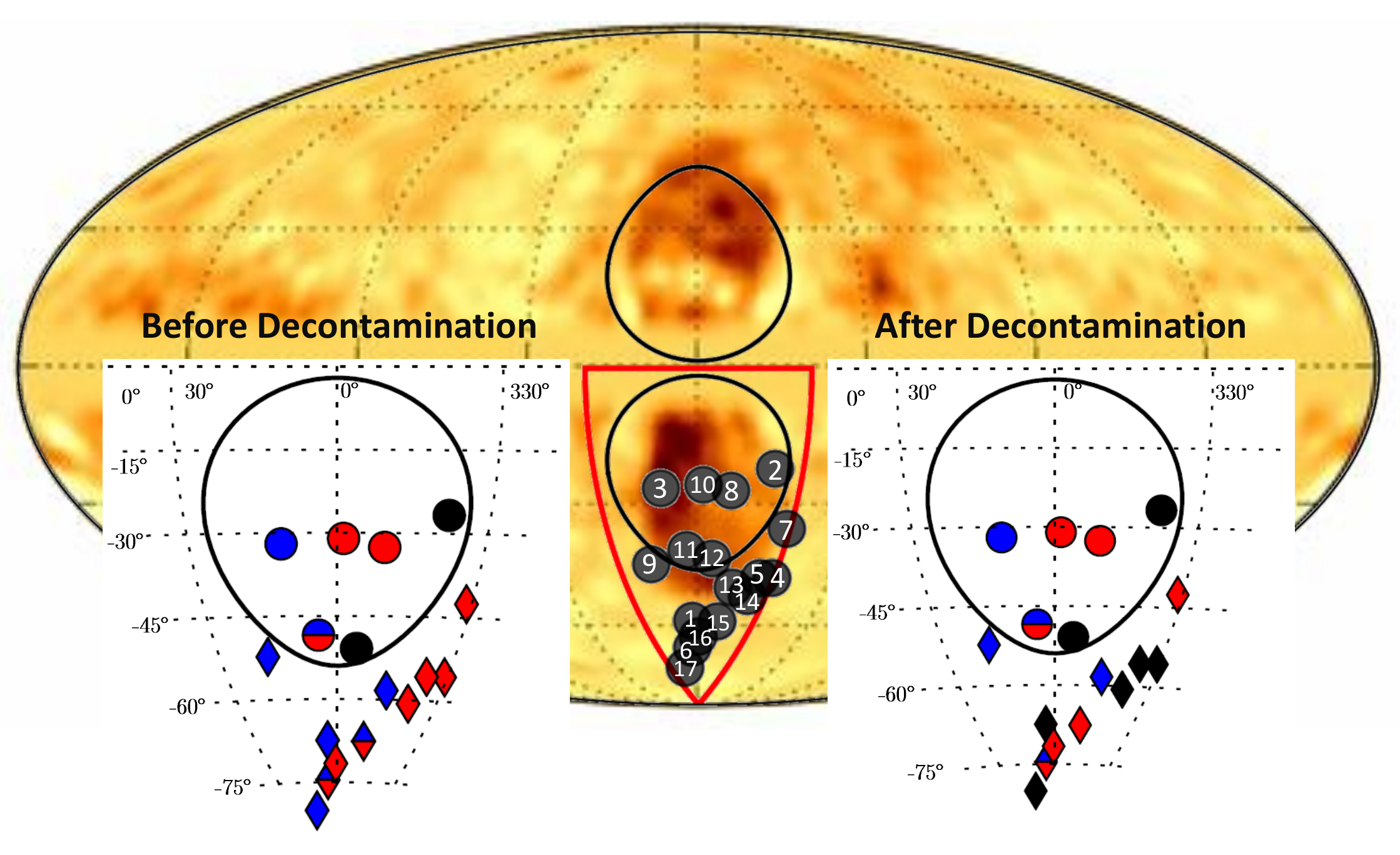}
    \caption{\textit{Fermi} all-sky map of gamma-ray residual emission in the 3--10 GeV range in Galactic coordinates. The Fermi Bubbles are shown in dark orange \citep[reproduced from][]{Ackermann14}. The approximate contour of the Fermi Bubbles is shown in black, while the search zone for the 17 sightlines is shown in red ($330^{\circ} < l < 30^{\circ}$ and $-90^{\circ} < b < 0^{\circ}$). The black circles denote the \hst/COS sightlines, and the numbers denote the ID number of the sightlines in Table~\ref{tab:qso}. Some of the circles' positions have been offset for visual representation. The gamma-ray emission at $|b| < 10\degr$ is contaminated by the Galactic disk emission, which has been replaced with the faint yellow band. The two inset panels show zoomed-in maps of the southern Fermi Bubble region color-coded by velocity of HVC components, before (left) and after (right) the removal of Magellanic contamination. The oval contours denote the approximate edge of the Bubble. Circles and diamonds denote sightlines inside and outside the southern Bubble, respectively. Red, blue and black symbols denote positive (redshifted), negative (blueshifted), and no HVCs. A red and blue split symbol indicates that both positive and negative HVCs are identified in that sightline. Seven sightlines show Magellanic Stream contamination.}
    \label{fig:FermiGamma}
\end{figure*}

As the home of the closest supermassive black hole, the Galactic Center (GC) is a fascinating laboratory for astronomers and physicists alike. \citet{Bland-Hawthorn03} detected the existence of two bipolar structures at the GC in hard X-rays; later on, \citet{Su2010} detected two giant diffuse lobes of plasma in gamma ray, called the Fermi Bubbles (shown in Figure~\ref{fig:FermiGamma}), while looking for signs of dark matter at the GC using the \textit{Fermi Gamma-ray Space Telescope}. These Bubbles coincided with the bipolar structures of \citet{Bland-Hawthorn03}. The Bubbles extend $\approx$55$\degr$ ($\approx$12 kpc) on either side of the GC  and show little variation in their surface brightness. 
The Fermi Bubbles are detected in emission in multiple wavelengths including hard X-rays \citep{Bland-Hawthorn03}, soft X-rays \citep{Kataoka13}, $\gamma$-ray \citep{Su2010}, K-band microwaves \citep{Dobler08} as the so-called \textit{Wilkinson Microwave Anisotropy Probe (WMAP)} haze, and polarized radio waves at 2.3 GHz \citep{Carretti13}. 
Their origin and nature is an active area of observational and theoretical research
\citep[see][and references therein]{Dobler2010, Su2010, Ackermann14, Yang2012, Yang2014, Crocker2011, Crocker2015}.


The Fermi Bubbles provide clues on the influence and importance of nuclear outflows on galactic evolution. 
Recently, our collaboration has made new contributions to the understanding of the Fermi Bubbles: we have used ultraviolet (UV) spectroscopy to study the gas kinematics, chemical abundances, and spatial extent of the cool gas clouds embedded inside the Fermi Bubbles.
Using the {\it Hubble Space Telescope} Cosmic Origins Spectrograph (\hst/COS), we have conducted UV-absorption spectroscopy of distant 
active galactic nuclei (AGN). \citet{Fox15} and \citet[hereafter B17]{Bordoloi17} studied the northern Fermi Bubble, by searching for the existence of any high-velocity cloud (HVC) absorption components whose kinematics indicated a possible origin in the Fermi Bubbles. HVCs are multi-phase clouds of interstellar gas that have velocities incompatible with the differential galactic rotation \citep{Wakker1997}, corresponding to LSR velocities $|v_{\rm LSR}|\geq90$\kms\ (more details in Section~\ref{sec:lsr}). The term HVC was traditionally used to refer to neutral 21 cm clouds, but we use it to refer to clouds of any ionization state (that may be predominantly ionized).

HVCs are important for understanding the role of gas accretion and outflow in Galactic evolution \citep{Putman2012}, as they represent a large fraction of the Galactic baryonic inflow rate \citep[e.g.][]{Richter2017}.
HVCs associated with the Fermi Bubbles can shed light
on the properties (e.g. mass, phase structure, and age) of the cool gas entrained in the nuclear wind.

In this paper, we study the {\it southern} Fermi Bubble using UV absorption spectra of 17 different sightlines in the southern galactic hemisphere, including directions both inside and outside the gamma-ray emission contours. The only other published UV analysis of a complete sightline through the southern Fermi Bubble is the recent study of \citet{Savage2017}, who examined the absorption properties of the sightline to the distant blue supergiant LS~4825. \citet{Keeney2006} also published an \hst/STIS spectrum of the PKS 2005-489 sightline, just outside the southern Fermi Bubble, and interpreted the absorption in terms of a nuclear wind. Several other UV-absorption studies have probed gas in the vicinity of the GC  and the inner galaxy \citep{Zech2008, Bowen2008, Wakker2012}.

Our paper is structured as follows.
In Section~\ref{sec:data} we describe our data sources and instruments. In Section~\ref{sec:method}, we describe our methodology to detect and measure HVC components in the absorption spectra of the Southern Fermi Bubble. 
We present the results in Section~\ref{sec:results} and discuss our inferences in Section~\ref{sec:discussion}. Finally, we summarize our findings in Section~\ref{sec:conclusion}.

\section{Observations and Data Handling}
\label{sec:data}

There are two principal datasets presented in this paper: (1) UV absorption-line spectra of background AGN acquired using \hst/COS, and (2) \hi\ 21-cm emission data from the Parkes Galactic All Sky Survey (GASS) \citep{McClure-Griffiths2009}. In addition, in Appendix~\ref{sec:saltsec} we present SALT/HRS optical spectra of \ion{Ca}{2} and \ion{Na}{1} absorption for two of our sightlines -- PKS 2155-304 and PDS\,456. The COS and GASS datasets are discussed in detail in Sections \ref{sec:cosdata} and ~\ref{sec:gassdata}, respectively, but first we describe how the sample was created.

\subsection{Sample Selection}
\label{sec:sample}
Our AGN sample was formed by identifying all \hst/COS datasets of AGN in the Southern Fermi Bubble region with spectra from the FUV/G130M grating. We define this region as $-90\degr\!<\!b\!<\!0\degr$ and $330\degr\!<\!l\!<\!30\degr$, to encompass directions passing through the Bubble and those passing nearby (see Figure~\ref{fig:FermiGamma}). The search region intentionally covers a region larger than the Bubble itself so that we can compare the properties of directions through the Bubble to those passing ourside. It extends down to $b=-90\degr$ since it is not known how far away from the Galactic plane the nuclear wind extends. 

Applying these criteria led to a sample of 17 sightlines, 
seven of which were observed in our own Fermi Bubble programs (\hst\ Program IDs 12936, PI=E. Jenkins, and 13448, PI=A. Fox). The remaining ten were taken from archival programs.
%
Table~\ref{tab:qso} lists these AGN with their galactic latitude and longitude, the \hst\ Program the data were taken under, and a flag indicating whether the sightline passes inside or outside the Bubble. Roughly a third (6/17) of the sightlines pass inside and the rest pass outside.

\begin{deluxetable*}{llcccc}[!ht]
\tablecaption{The Sample: UV-bright AGN observed with \hst/COS in the Southern Fermi Bubble Region\tm{a}\label{tab:qso}}
\tabletypesize{\scriptsize}
\tablecolumns{3}
\tablehead{\colhead{ID \tm{b}} & \colhead{Sightline} & \colhead{$l$\tm{c}} & \colhead{$b$\tm{c}} &\colhead{Prog. ID\tm{d}} & \colhead{Location\tm{e}}\\ 
\colhead{} & \colhead{} & \colhead{($^{\circ}$)} & \colhead{($^{\circ}$)} & \colhead{} & \colhead{}}

\startdata
1 & \object{CTS487} & 5.54 & $-$69.44       & 13448 & Out\\
2 & \object{ESO141-G55} & 338.18 & $-$26.71 & 12936 & In (Border)\\
3 & \object{ESO462-G09} & 11.33 & $-$31.95  & 13448 & In\\
4 & \object{HE2258-5524} & 330.72 & $-$55.67& 13444 & Out\\
5 & \object{HE2259-5524} & 330.64 & $-$55.72& 13444 & Out\\ 
6 & \object{HE2332-3556} & 0.59 & $-$71.59  & 13444 & Out\\ 
7 & \object{IRAS F21325-6237} & 331.14 & $-$42.52 & 12936 & Out\\ 
8 & \object{PKS2005-489} & 350.37 & $-$32.60& 11520 & In\\
9 & \object{PKS2155-304} & 17.73 & $-$52.25 & 12038 & Out \\ 
10 & \object{RBS1666} & 358.73 & $-$31.00    & 13448 & In\\
11 & \object{RBS1768} & 4.51 & $-$48.46      & 12936 & In (Border)\\ 
12 & \object{RBS1795} & 355.18 & $-$50.86    & 11541 & In (Border)\\ 
13 & \object{RBS1892} & 345.9 & $-$58.37     & 12604 & Out\\
14 & \object{RBS1897} & 338.51 & $-$56.63    & 11686 & Out\\
15 & \object{RBS2000} & 350.20 & $-$67.58    & 13448 & Out\\
16 & \object{RBS2023} & 0.61 & $-$71.62      & 13444 & Out\\
17 & \object{RBS2070} & 12.84 & $-$78.04     & 12864 & Out\\
\enddata
\tn{a}{The Southern Fermi Bubble region is defined here as $-90\degr\!<\!b\!<\!0\degr$ and $330\degr\!<\!l\!<\!30\degr$, larger than the Bubble itself.}
\tn{b}{ID number shown in Figure~\ref{fig:FermiGamma}.}
\tn{c}{Galactic longitude and latitude.}
\tn{d}{\hst\ Program ID of dataset analyzed.}
\tn{e}{Whether the sightline passes inside or outside the gamma-ray emission contours from the Bubble (Figure 1). Three of the inside sightlines are close to the boundary.}
\end{deluxetable*}

\subsection{Cosmic Origins Spectrograph (COS)}
\label{sec:cosdata}

COS is a moderate-resolution spectrograph installed on \hst\ in 2009. We use data from both medium-resolution far-ultraviolet (FUV) gratings (G130M and G160M) for directions where they exist, though all directions have G130M spectra. The G130M grating covers $\approx$1150--1450\AA\, and G160M covers $\approx$1450--1750\AA, depending on the central wavelength setting.
Together these gratings cover the wavelength range of $\approx$1150--1700 {\AA} (depending on the central wavelength setting used, which varies from case to case), and the resolving power of both gratings (at the FUV detector lifetime positions in use when the data were taken) was between 16000--21000\footnote{\url{http://www.stsci.edu/hst/cos/design/gratings}}. The final science spectra have S/N $\approx$ 12--20 per resolution element, a velocity resolution of $\approx$ 20 \kms\ (FWHM), and an absolute velocity uncertainty scale of $\approx$ 5 \kms\ \citep{Fox2017}. 

The archival data were obtained from the Mikulski Archive for Space Telescopes (MAST)\footnote{\url{https://archive.stsci.edu}}. All the COS data (new plus archival) were reduced using our own custom-designed software, which cross-correlates exposures to align low-ion interstellar lines in velocity space before co-addition and aligns the ISM absorption lines with 21-cm emission, and makes sure that lines in intergalactic absorption systems are properly aligned \citep[for full details see][]{Wakker2015}. For visualization in Appendix~\ref{sec:voigtprofiles}, the spectral data were rebinned by 5 pixels (which gives two binned pixels per resolution element). However, for the measurements and analysis described in Section~\ref{sec:method} we used the unbinned data.

\subsection{The Parkes Galactic All Sky Survey (GASS)}
\label{sec:gassdata}

To determine the \hi\ column densities (or upper limits on them) in each HVC in our survey, we use the GASS survey \citep{McClure-Griffiths2009}. GASS is a survey of \ion{H}{1} 21-cm emission of the entire southern sky up to declination of +1$\degr$. It covers LSR velocities between $-$400 \kms\ and +500 \kms\ and provides spectra with 16$\arcmin$ angular resolution, 0.82 \kms\ spectral resolution, and an RMS sensitivity of 57 mK. 


\section{Methodology: COS Analysis}
\label{sec:method}

For the absorption-line analysis, we identified 11 spectral lines, chosen as the strongest absorption lines in the COS FUV bandpass. Table~\ref{tab:ions} lists these 11 lines, their wavelengths, and their oscillator strengths \citep[$f$-values; taken from][]{Morton2003}. We checked the atomic data against the recent compilation of \citet{Cashman2017}.
We searched the 17 sightlines in our sample for absorption-line HVCs in these spectral lines. The search was conducted over the LSR velocity interval $-$
500 to 500 \kms. HVCs are defined as components with $|v_{\rm LSR}|>90$ \kms. 
Contamination by intergalactic absorbers was ruled out since we conducted a line-by-line identification of each COS spectrum, and checked that no IGM lines blended with our HVC metal lines. 

We only included HVCs in our sample if they were detected in two or more transitions. This ensures that detector artifacts and unidentified blends do not contaminate the sample. The single exception to this rule is the HVC at
$-$270 \kms\ toward PKS 2155-304, which is only detected at high significance in \ion{Si}{3} 1206, because its reality seems confirmed by a weak (but low significance) counterpart in \ion{C}{2} 1334.

\begin{deluxetable}{lll}[!hb]
\tablecaption{UV lines analyzed\label{tab:ions}\tm{a}}
\tabletypesize{\scriptsize}
\tablecolumns{3}
\tablehead{\colhead{Wavelength} & \colhead{Transition} & \colhead{$f$-value}\\ 
\colhead{(\AA)} & \colhead{} & \colhead{}}

\startdata
1334.5323 & \ion{C}{2} 1334 & 0.1278\\
1670.7874 & \ion{Al}{2} 1670 & 1.8800\\
1260.4221 & \ion{Si}{2} 1260 & 1.0070\\
1193.2897 & \ion{Si}{2} 1193 & 0.49910\\
1190.4158 & \ion{Si}{2} 1190 & 0.25020\\
1526.7066 & \ion{Si}{2} 1526 & 0.1270\\
1206.500 & \ion{Si}{3} 1206 & 1.660\\
1548.195 & \ion{C}{4} 1548 & 0.19080\\
1550.770 & \ion{C}{4} 1550 & 0.095220\\
1393.755 & \ion{Si}{4} 1393 & 0.5280\\
1402.770 & \ion{Si}{4} 1402 & 0.2620\\
\enddata
\tn{a}{Atomic data from \citet{Morton2003}.}
\end{deluxetable}



\subsection{Converting to LSR Velocity}
\label{sec:lsr}
The wavelength arrays produced by the COS data reduction pipeline are given in the heliocentric reference frame.
By convention, Galactic absorption spectra are plotted as flux versus velocity in the local standard of rest (LSR), $v_{\rm LSR}$, which accounts for the peculiar motion of the Sun. The LSR is defined as a point in space that is moving on a circular orbit at the Solar radius about the GC  \citep{Huang15}. We converted heliocentric wavelengths and velocities to the LSR frame by applying an offset where:





\begin{equation}
     v_{\rm LSR} = v_{\rm helio} + v_{\rm corr}
\end{equation}
and where $v_{\rm corr}$ in \kms\ is given by:
\begin{equation}
    v_{\rm corr} = 9.0 \cos{l} \cos{b} + 12.0 \sin{l} \cos{b} + 7.0 \sin{b}     
\end{equation}




\subsection{Voigt-Profile Fitting}
\label{sec:vpfit}

To fit the continuum, we examined each absorption line in velocity space in a region extending $\pm$500 \kms\ around the line center, and performed a polynomial fit of degree 1 in the unabsorbed part of the spectra using the Python package {\tt Numpy} \citep{Walt2011}. We then used the polynomial fit to normalize the flux. 

Afterwards, we used the software packages {\tt RDGEN}\footnote{\url{http://www.ast.cam.ac.uk/$\sim$rfc/rdgen9.5.pdf}} \citep{RDGEN2014} and {\tt VPFIT}\footnote{\url{http://www.ast.cam.ac.uk/$\sim$rfc/vpfit.html}} \citep{VPFIT2014} to fit observed HVC absorptions in the normalized flux data with Voigt components. {\tt RDGEN} is an interactive program associated with {\tt VPFIT} which allows users to determine initial estimates for absorption features. These estimates are then formatted as outputs by {\tt RDGEN}, which are then used as inputs for {\tt VPFIT}. {\tt VPFIT} utilizes Voigt profile 
fitting algorithms to model the HVCs. The Voigt profiles use three parameters to describe each absorption component: (1) velocity centroid, (2) Doppler parameter, and (3) column density. 
{\tt VPFIT} allows multiple components to be fit simultaneously to multiple absorption lines. It works by minimizing $\chi^{2}$ in an iterative process using Gauss-Newton type parameter updates  \citep[for more discussion, see Appendix A-2 of][]{Murphy2003}. 



\subsection{Identification of Galactic Center Components}
\label{sec:identification}

Once the absorption-line HVCs were measured, we checked to make sure that the same HVC has been detected in at least two different ions. This serves as a check against falsely considering blends or interlopers as positives.
After this step, we identified which HVCs arise in the GC and which have alternative origins. There are two chief alternative origins that we consider: the foreground Galactic disk and the Magellanic Stream. Absorption components due to either of these sources need to be identified and removed. 

\subsubsection{ISM Decontamination}
\label{sec:ism-blend}

We use a model of a co-rotating cylindrical ISM to determine, for each direction, the minimum and maximum LSR velocity expected for gas in co-rotation with the Galactic disk \citep[following][]{Savage1987}. Any absorption found within this velocity interval  $\Delta v$(Gal) is likely produced by foreground ISM absorption and cannot be attributed to the Fermi Bubbles. Table~\ref{tab:vpfitresults} shows the velocity range of co-rotation computed for each direction in our sample. The size of $\Delta v$(Gal) is a strong function of Galactic longitude: directions close to the GC  (small $|l|$) have very small values of $\Delta v$(Gal) because the line-of-sight in these directions is almost perpendicular to the direction of co-rotation. In all cases in our sample, $\Delta v$(Gal)$<$ 20 \kms, and so none of the observed HVCs (which by definition have $|v_{\rm LSR}|>90$ \kms) are due to Galactic co-rotation.

\subsubsection{Magellanic Stream Decontamination}
\label{sec:magstream}

The Magellanic Stream is an extended arc of gas originating from the Magellanic Clouds and covering a significant fraction of the southern sky.
\citep{Putman2003, Nidever2008, donghia2016}.
Sightlines passing through both the Southern Fermi Bubble and the Magellanic Stream will contain multiple HVCs in their absorption spectra. 
In order to identify Magellanic components, we made use of the velocity map of the Magellanic Stream presented in \citet{Fox2014}, which is based on the \hi\ velocity field of \citet{Nidever2008}, and extended to include the ionized regions of the Stream.
By comparing the velocity of each observed HVCs in our sample with the expected Magellanic velocity in that direction from the published map, we can determine whether the HVC is likely to be Magellanic. 


The \citet{Fox2014} velocity map is in the Magellanic Stream Coordinate System, with Magellanic longitude and latitude $\lambda_{\rm MS}$ and $\beta_{\rm MS}$. The conversion between $l,b$ and $\lambda_{\rm MS}$ and $\beta_{\rm MS}$ is 
given by 
\footnote{\url{http://www.atnf.csiro.au/people/Tobias.Westmeier/tools\_coords.php}}:

\begin{equation}
    \tan (\lambda_{\rm MS} - \lambda_{\rm MS,0}) = \frac{\sin b \sin \epsilon + \cos b \cos \epsilon \sin (l-l_{0})}{\cos b \cos (l-l_{0})}
\end{equation}

\begin{equation}
    \sin \beta_{\rm MS} = \sin b \cos \epsilon - \cos b \sin \epsilon \sin (l-l_{0})
\end{equation}

where $\lambda_{\rm MS,0}$=278$^{\circ}$.5, $\lambda_{\rm MS, 0} \simeq$ 32$^{\circ}$.8610, and $\epsilon$=97$^{\circ}$.5.

The effect of the removal of Magellanic components from our sample is shown in in the insets of Figure~\ref{fig:FermiGamma}. 
Out of the 22 HVCs identified in the full sample, 7 can be ascribed a Magellanic origin, leaving 15 as the candidates for the GC outflow.


\section{Results}
\label{sec:results}

\subsection{Example Spectra: RBS 1768 and RBS 2000}
\label{sec:examples}
\begin{figure*}[!ht]
    \epsscale{0.85}
    \plotone{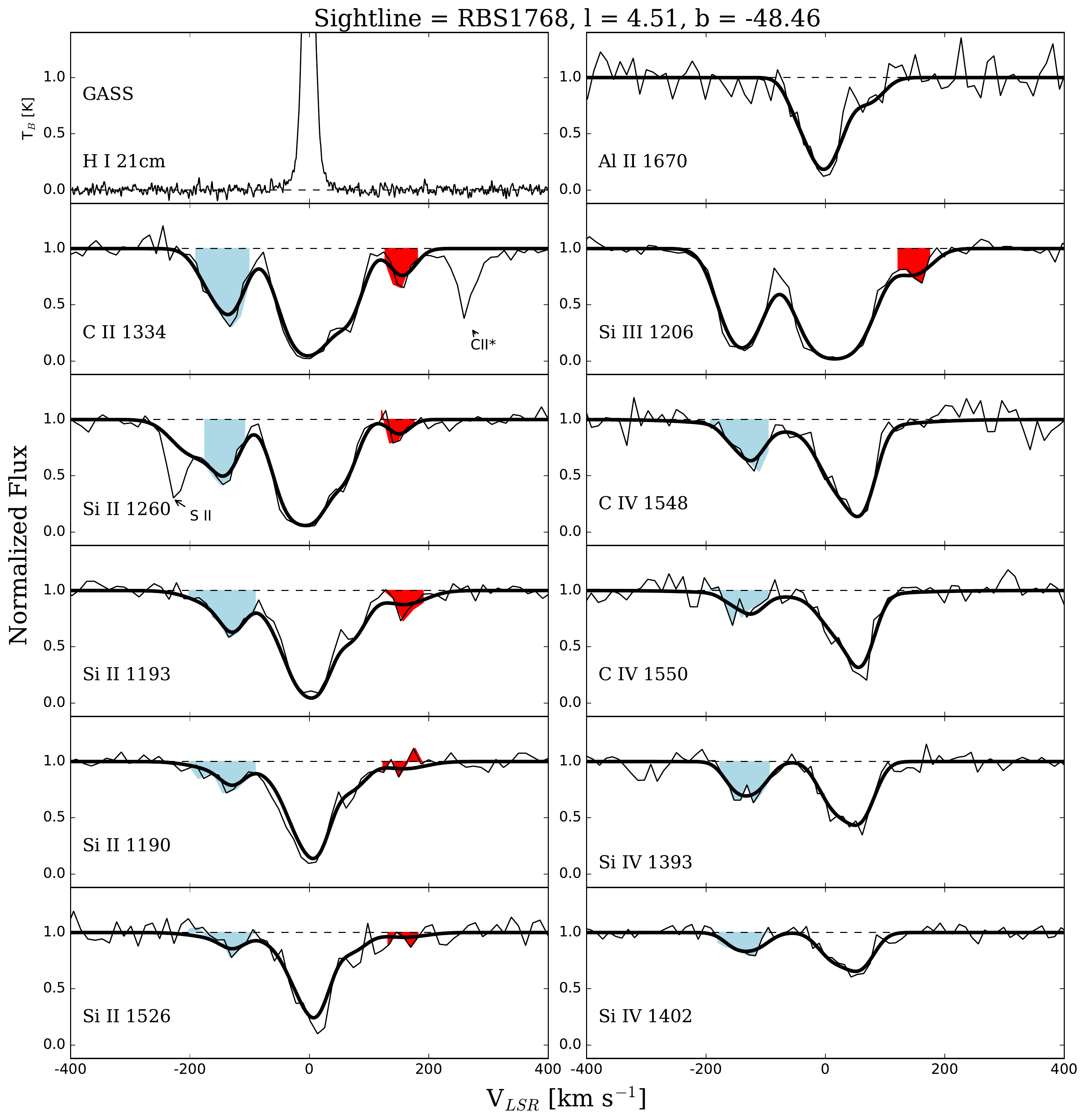}
    \caption{\hst/COS UV absorption spectra of the AGN RBS 1768 (a sightline passing through the Bubble; ID 11 in Figure 1). Normalized flux is plotted against LSR velocity for a range of UV metal absorption lines. The VPFIT model is shown as the red line. The \hi\ 21 cm profile from GASS is included in the top-left panel.}
    \label{fig:1768}
\end{figure*}

\begin{figure*}[!ht]
    \epsscale{0.85}
    \plotone{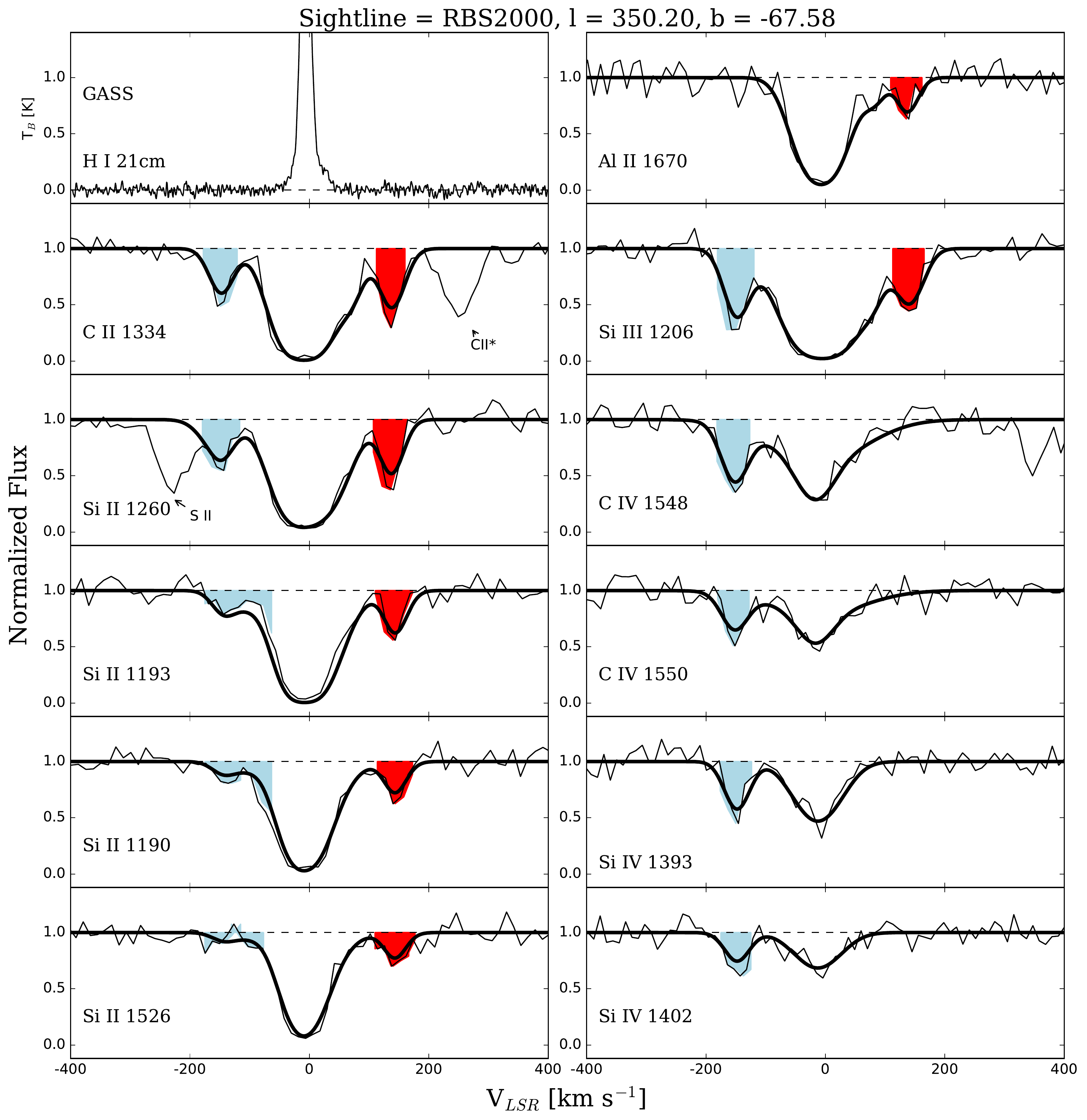}
    \caption{\hst/COS UV absorption spectra of the AGN RBS 2000 (a sightline passing outside the Bubble; ID15 in Figure 1). Normalized flux is plotted against LSR velocity for a range of UV metal absorption lines. The VPFIT model is shown as the red line. The \hi\ 21 cm profile from GASS is included in the top-left panel.}
    \label{fig:2000}
\end{figure*}

To illustrate the quality of the data and the lines used for measurement, we present the \hst/COS spectra for two representative sightlines, RBS 1768 and RBS 2000 in Figures~\ref{fig:1768} and \ref{fig:2000}. The {\tt VPFIT} models are overplotted in red. These two cases were chosen to include one direction (RBS 1768) passing through the Southern Fermi Bubble, and one (RBS 2000) passing outside. LSR velocity stacks showing the absorption profiles for the 15 other directions are included in the figure set Appendix~\ref{sec:voigtprofiles}. 

Both the RBS 1768 and RBS 2000 spectra show a similar ``twin-component" HVC absorption profile, with a combination of negative and positive HVC components: one at $v_{\rm LSR}$=$-$150 \kms\ and the other at $v_{\rm LSR}$=+160 \kms. 
The twin-component HVC structure was seen by \citet{Fox15} toward the QSO PDS456 lying close to the GC, and interpreted in terms of a signature of a biconical outflow. We also noticed a blend in \ion{Si}{2} 1190 in both of these sightlines at $-$100 \kms. It is notable that both the inside (RBS1768) and outside (RBS2000) sightlines show the twin component structure. While we attribute both of the components of RBS 1768 to the Southern Bubble, we found that the negative component of RBS 2000 is due to the Magellanic Stream.

Because the HVCs in many sightlines have a complex velocity structure, our Voigt-profile models should be viewed as an approximation. The complex velocity structure may arise due to a combination of broad and narrow features and/or contamination with blends. As a result, our best modeled profiles are underfits in a few cases, e.g. the redshifted HVC in \ion{Si}{2} 1260 toward RBS 1768 and RBS 2000 (see Figures \ref{fig:1768} and \ref{fig:2000}). Our approach was to optimize the fits for the HVC absorption profiles, sometimes at the expense of the fits to the low-velocity absorption, which is of less scientific interest for this paper.

\subsection{\ion{H}{1} 21 cm emission non-detections}
\label{sec:h1limits}
None of the HVCs in our survey is detected in \hi\ 21 cm emission in the GASS data, which has a 3$\sigma$ limiting column density sensitivity log $N$(\hi)=18.20 \citep{McClure-Griffiths2009}. This lack of \hi\ together with the detection of many low-ion and high-ion metal lines indicates a high level of ionization in the GC  HVCs. This is consistent with what is seen in the northern hemisphere \citep{Fox15, Bordoloi17}.

\subsection{Incidence of HVCs}
\label{sec:idenhvc}
A straightforward observational constraint on the presence of gas within the Fermi Bubbles is the incidence of HVC absorption, equivalent to the gas covering fraction. 
We found that HVCs are detected in 15 out of 17 sightlines in the sample. 
After we removed the Magellanic Stream HVCs as described in Section~\ref{sec:magstream},
four out of six sightlines passing through the Southern Bubble show HVCs, whereas six out of eleven sightlines outside the Bubble show HVCs. 
To investigate whether the distributions of HVCs inside and outside the Southern Bubble are the same, we constructed 2x2 contingency tables and performed the adjusted Pearson's $\chi^2$-test, where the adjustment refers to the inclusion of Yates' correction for continuity \citep{Yates1934}. This correction is important for small sample numbers in each element of the contingency table. Without the correction the traditional Pearson's $\chi^2$-test may overestimate the statistical significance of any result drawn from a small sample. Table~\ref{tab:conttable} shows the 2x2 contingency table we used for blueshifted HVCs.

\begin{deluxetable}{rrr|r}[!h]
\tablecaption{2x2 Contingency Table for Blueshifted HVCs.
\label{tab:conttable}}
\tabletypesize{\scriptsize}
\tablecolumns{3}
\tablehead{
\colhead{Location} & \colhead{No. of sightlines} & \colhead{No. of sightlines} &  \colhead{Total}\\
& \colhead{with HVCs} & \colhead{ without HVCs} &}
\startdata
Inside & 2 & 4 & 6\\
Outside & 3 & 8 & 11\\ 
\hline
Total & 5 & 12 & 17\\
\enddata
\end{deluxetable}

When we apply the adjusted Pearson's $\chi^2$ test, we find that for redshifted HVCs, blueshifted HVCs, and all HVCs, the p-values obtained are 0.98, 0.77, and 0.98 respectively\footnote{For comparison, if we conduct a conventional Pearson's $\chi^2$ test without Yates' correction, the corresponding p-values are 0.59, 0.79, and 0.63, respectively.}. \textit{Thus, we are unable to reject the null hypothesis that the distribution of HVCs inside and outside the Southern Bubble are drawn from the same population}. This is an interesting deviation from the findings of B17 on the Northern Fermi Bubble, who report a clear and significant enhancement in the gas covering fraction inside the Bubble.
We suspect that this is partly due to
small-number statistics in the south (especially given that we have sub-divided the sample into negative- and positive-velocity HVCs), as well as the fact that in general, there are more HVCs surrounding the southern bubble than surrounding the northern bubble (many of which may be unrelated to the Fermi Bubbles), so the background is higher in the south, even after the removal of the Magellanic Stream.

In order to address the issue of small number statistics, it is useful to consider the two Fermi Bubbles together as a pair. To do this, we combine our data on the southern bubble with the B17 data on the northern bubble, and redo the adjusted Pearson's $\chi^2$-test to determine whether the distributions of HVCs inside and outside the bubbles are the same. When we do this, we find that
for redshifted, blueshifted, and all HVC cases, the p-values are 0.004, 0.97, and 0.02 respectively. \textit{Thus, when we combine data from the Northern and Southern Bubbles, we can reject the null hypothesis that the incidence of HVCs inside the Bubbles are the same as the incidence of HVCs outside the Bubbles. This shows that overall, there is a statistical enhancement in UV-absorbing gas in directions passing inside the Bubbles versus those passing outside, but this signal is dominated by the northern bubble.}

A summary of these findings on HVC detection rates is shown in Table~\ref{tab:allskytab}. The results from B17 on the northern Fermi Bubble are included for comparison. For comparison, all-sky HVC covering fractions for UV absorption are $\approx$60--80\%, depending on the line used \citep{Shull2009, Lehner2012, Richter2017}.

\begin{deluxetable}{lrrrr}[!h]
\tablecaption{Comparison of HVC incidence inside and outside the Southern\tm{1} and Northern\tm{2} Fermi Bubbles.
\label{tab:allskytab}}
\tabletypesize{\scriptsize}
\tablecolumns{3}
\tablehead{
\cutinhead{Southern Bubble} \\
\colhead{Location} & \colhead{No. of} & \colhead{No. of} &  \colhead{Positive} & \colhead{Negative} \\
& Sightlines & HVCs & HVCs\tm{3} & HVCs\tm{3}}
\startdata
Inside & 6 & 7 & 3 (50$\pm$19\%) & 4 (64$\pm$18\%)\\
\textit{... border\tm{4}} & \textit{3} & \textit{3} & \textit{1 (38$\pm$24\%)} & \textit{2 (62$\pm$24\%)}\\
Outside & 11 & 8 & 4 (50$\pm$17\%) & 4 (50$\pm$17\%)\\ 
\hline
\textbf{Total} & \textbf{17} & \textbf{15} & \textbf{7 (42$\pm$12\%)} & \textbf{8 (47$\pm$12\%)}\\
\cutinhead{Northern Bubble\tm{2}}
Inside & 5 & 6 & 1 (25$\pm$17\%) & 5 (92$\pm$8\%) \\
Border & 8 & 4 & 0 (6$\pm$6\%) & 4 (50$\pm$17\%) \\
Outside & 34 & 9 & 4 (13$\pm$6\%) & 5 (16$\pm$6\%) \\
\hline
\textbf{Total} & \textbf{47} & \textbf{19} & \textbf{5 (11$\pm$5\%)} & \textbf{14 (30$\pm$7\%)}\\
\hline
\hline
\textbf{Grand Total} & \textbf{64} & \textbf{34} & \textbf{13 (21$\pm$5\%)} & \textbf{21 (33$\pm$6\%)}\\
\enddata
\tn{1}{Magellanic HVCs have been removed from these statistics.}
\tn{2}{From B17.}
\tn{3}{These two columns show the number of sightlines with positive and negative HVCs. The percentages denote the ratio of number of HVCs to the number of sightlines. The error bars have been calculated using the Wilson score interval \citep{Wilson1927} to estimate the underlying binomial hit rates.}
\tn{4}{Treated as a sub-class of the class Inside.}
\end{deluxetable}

\subsection{Velocity vs Galactic Latitude}
\label{sec:vglat}

\begin{figure}[!ht]
    \epsscale{1.3}
    \plotone{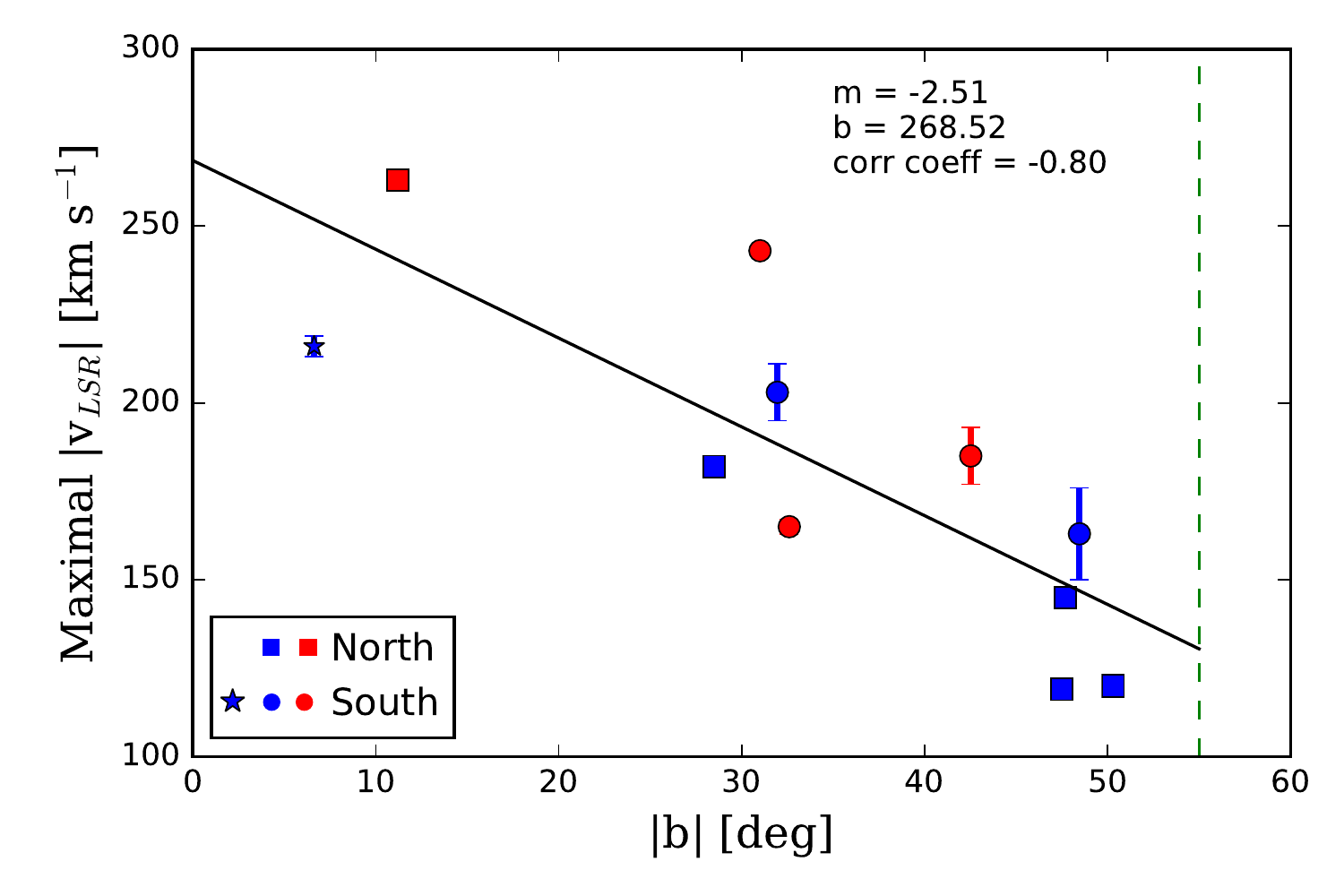}
    \caption{Maximal absolute LSR velocity as a function of Galactic latitude for HVCs in the Northern and the Southern Fermi Bubbles region. The data for the Northern Bubble (squares) are from B17. For the Southern Bubble (circles), \ion{Si}{3} is used whenever available; otherwise, \ion{C}{2}, \ion{C}{4}, \ion{Si}{4}, and \ion{Al}{2} were used. The star denotes the measurement from \citet{Savage2017} for a low-latitude blue supergiant star in the galactic south. Blue and red symbols represent blueshifted and redshifted HVCs, respectively. The dashed line represents the approximate boundary of the Fermi Bubbles.}
    \label{fig:velvglat}
\end{figure}

In order to investigate the wind radial profile, we plot the maximal absolute $v_{\rm LSR}$ of the identified HVCs as a function of absolute galactic latitude $|b|$ in Figure~\ref{fig:velvglat}. The maximal selection means we plot the highest absolute velocity of any HVC observed in each sightline, which is important since some sightlines show multiple components.
$|v_{\rm LSR}|$ is measured using \ion{Si}{3} 1206 whenever possible, since that is the strongest line available in our data. Only directions at $|b|<55\degr$ are included on the plot since we are exploring the HVC velocity profile \emph{within} the Bubble.

In Figure~\ref{fig:velvglat} a clear negative slope is observed, indicating that HVCs closer to the GC  are moving faster compared to those at higher latitude. The correlation coefficient between the absolute galactic latitude and maximal absolute LSR velocity is $-$0.80, indicating a strong downhill linear relationship; $|v_{\rm LSR}|$ drops from $\approx$216 \kms\ at $|b|$=6\degr\ to $\approx$120 \kms\ at $|b|$=50\degr. This mirrors the same finding seen in the northern hemisphere (B17). Indeed, we plot data from both studies and notice that the slope and normalization of the northern and southern data points are similar. This is consistent with expectations for a decelerating outflow expanding into both northern and southern hemispheres.

\subsection{Ionization vs Galactic Latitude}
\label{sec:iglat}





\begin{figure*}[!ht]
\gridline{
    \fig{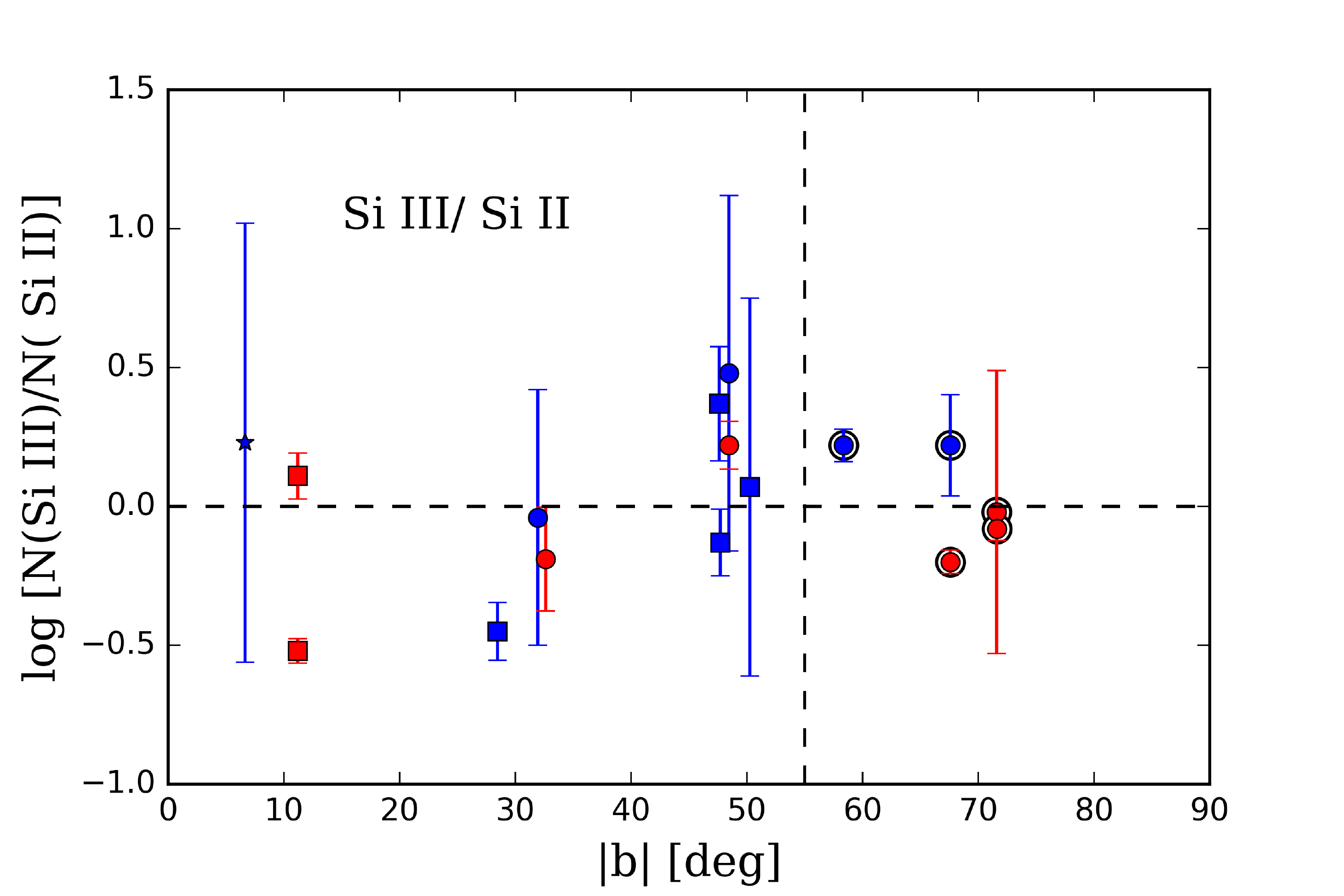}{0.5\textwidth}{}
    \fig{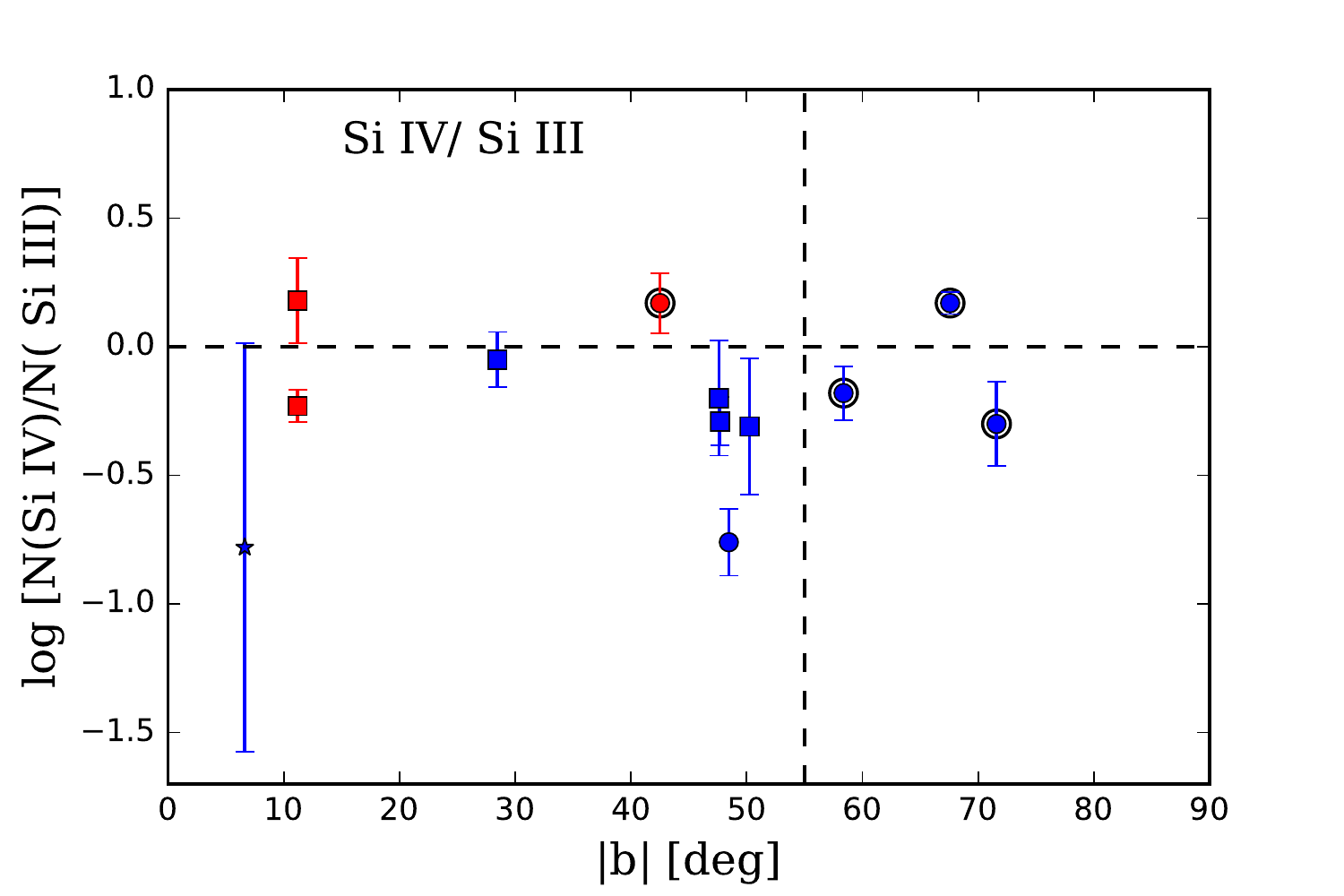}{0.5\textwidth}{}}
\gridline{
    \fig{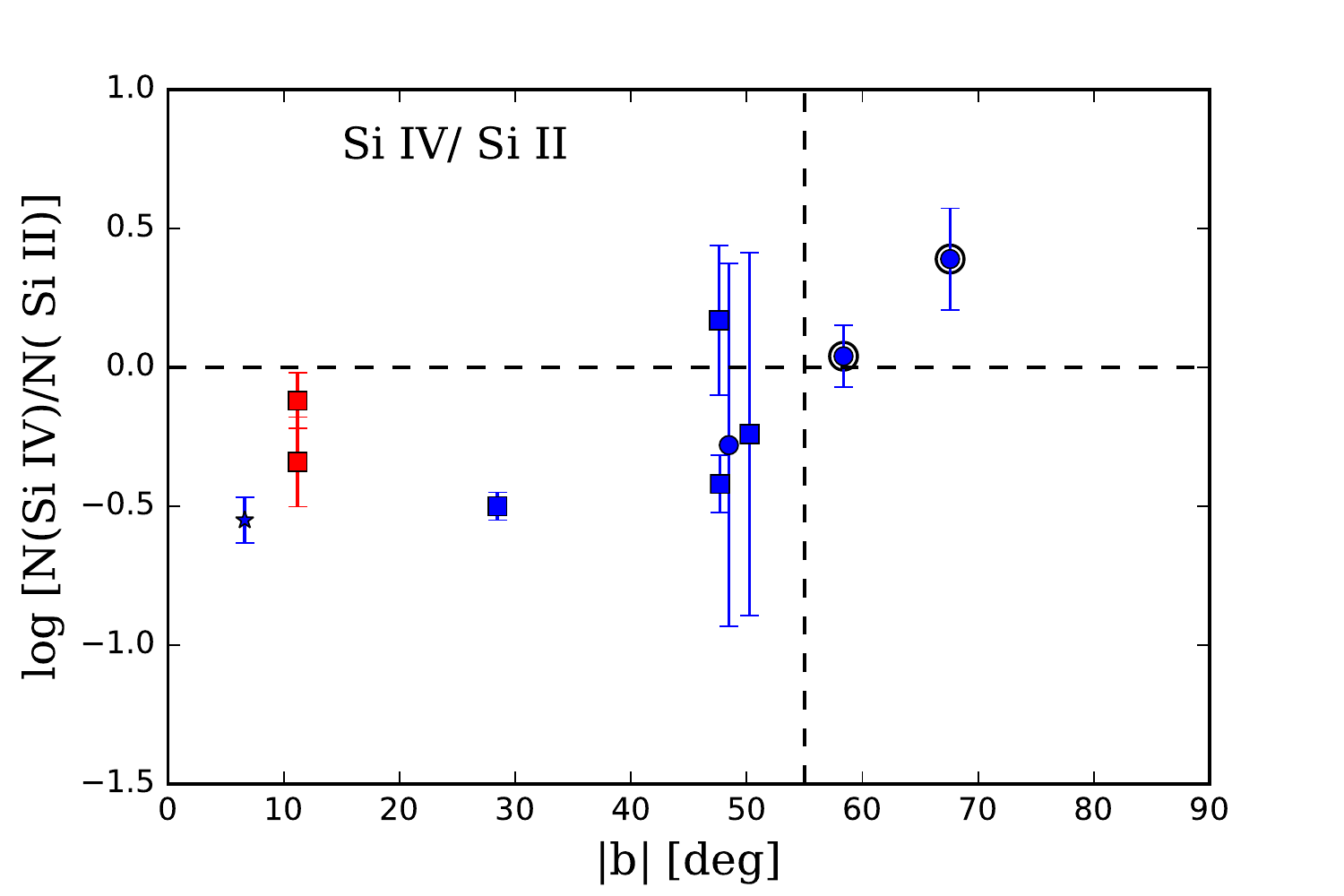}{0.5\textwidth}{}}
    \caption{Column-density ratios (a) \ion{Si}{3}/\ion{Si}{2}, (b) \ion{Si}{4}/\ion{Si}{3}, and (c) \ion{Si}{4}/\ion{Si}{2} in the GC  HVCs as a function of absolute galactic latitude $|b|$. Data from both the northern \citep[squares;][]{Bordoloi17} and the southern (circles; this study) Fermi Bubbles are included. The star denotes the measurement from \citet{Savage2017} for a low-latitude blue supergiant star in the Galactic south. Blue and red represents blueshifted and redshifted HVCs respectively, while the white edge denotes sightlines outside the Fermi Bubbles region. The vertical dashed line represents the galactic latitude boundary of the Fermi Bubbles, and the horizontal line represent the zero-level (equal amounts of each ion).}
    \label{fig:ionratios}
\end{figure*}

We use the column density ratios $N$(\ion{Si}{3})/$N$(\ion{Si}{2}), $N$(\ion{Si}{4})/$N$(\ion{Si}{3}) and $N$(\ion{Si}{4})/$N$(\ion{Si}{2}) to diagnose the ionization level in the HVCs. Using the ratio of two species of the same chemical element has the advantage that the ratio is independent of the absolute abundance of that element, and independent of dust depletion. 
\ion{Si}{2} is often classified as a low ion, \ion{Si}{3} as an intermediate ion, and \ion{Si}{4} as a high ion, so the ratios among these three ions gauges the relative amounts of low-ion, intermediate-ion, and high-ion gas.
The ratios are plotted as a function of absolute galactic latitude $|b|$ in Figure~\ref{fig:ionratios}, which includes data from the northern hemisphere from B17. 

None of the three plots in Figure~\ref{fig:ionratios} show any obvious trend.
We performed correlation analyses for each of the three ratios. The $N$(\ion{Si}{3})/$N$(\ion{Si}{2}), $N$(\ion{Si}{4})/$N$(\ion{Si}{3}) and $N$(\ion{Si}{4})/$N$(\ion{Si}{2}) plots have correlation coefficients $-$0.20, $-$0.14 and 0.35, and slopes $-$0.002, $-$0.002 and 0.004 respectively. Thus, there is barely any trend of significance of ion ratios as a function of absolute galactic latitude. In a cooling wind scenario, we expected to see higher ionization at lower galactic latitude because low latitude sightlines are closer to the energetic central engine. Thus, the flatness of all the three ion species is an interesting deviation from our expectation. 

\section{Discussion}
\label{sec:discussion}

\subsection{Comparison with Previous UV Absorption Studies}
\label{sec:o6comp}
Several Galactic Center sightlines, including five in our COS sample, have been studied in UV absorption previously with other spectrographs, namely the Space Telescope Imaging Spectrograph (STIS) or the Goddard High Resolution Spectograph (GHRS) onboard \hst, or the \emph{Far-Ultraviolet Spectroscopic Explorer (FUSE)}. We choose not to combine these earlier data with our COS sample, for two reasons. First, they have different properties in terms of spectral resolution and sensitivity, and second, many of these sightlines are already in our COS sample. Instead, we now briefly discuss the connection between our results and these earlier analyses.

\begin{itemize}
\item \emph{FUSE} surveyed the Galactic halo in \ion{O}{6} absorption  in 100 sightlines \citep{Savage2003, Wakker2003, Sembach2003}. Three of these sightlines are in our sample: PKS 2005-489, ESO 141-G55, and PKS 2155-304. 
Among all 100 sightlines in this \emph{FUSE} sample, PKS 2005-489 shows the \emph{highest} low-velocity ($|v_{\rm LSR}|<100$ \kms) \ion{O}{6} column density, with $\log$ N(\ion{O}{6}) $= 14.78 \pm 0.02$; the strength of this line is likely related to nuclear activity. In our COS data we find an HVC in \ion{Si}{2} and \ion{Si}{3} at 164 \kms, which may also be related to the GC wind.

\item \citet{Keeney2006} observed PKS 2005-489 using \textit{FUSE} and STIS, detecting two HVCs, one at $- 105\pm 12$ \kms\ in \ion{N}{5} and \ion{O}{6} and one at $+168 \pm 10$ \kms\ in \ion{Si}{2} and \ion{O}{6}.
In our COS data we detected both the blueshifted and the redshifted HVCs in \ion{Si}{2} and \ion{Si}{3}.

\item \citet{Bonamente2004} used \emph{FUSE} to observe to observe the IRAS F21325-6237 
and detected two \ion{O}{6} absorption features 
at $v_{\rm LSR}=27 \pm 3$ \kms\ and $175 \pm 6$ \kms. We detect the latter feature in the COS data in \ion{C}{4}, \ion{Si}{3}, and \ion{Si}{4}, with $v_{\rm LSR}$ $178 \pm 1$ \kms, $185 \pm 8$ \kms, and $170 \pm 3$ \kms, respectively.

\item \citet{Sembach1999} observed ESO 141-G55 with GHRS and with ORFEUS II and report 
strong \ion{Si}{4}, \ion{C}{4}, \ion{N}{5} and \ion{O}{6} absorption, 
including a weak HVC at $+120$ \kms\ in \ion{Si}{4} and \ion{C}{4}, which we do not detect in our COS dataset; this may be related to S/N.
\end{itemize}
Overall, our COS survey largely confirms the presence of the absorbers seen in these previous studies, and provides additional information on their ionization state and composition. We have also expanded the sample to a much larger size.

\subsection{Comparison with \ion{H}{1} 21-cm Observations}
\label{sec:h1ref}

\citet{McClure-Griffiths2013} studied the $-5\degr \leq l \leq +5\degr$ and $-5\degr \leq b \leq +5\degr$ region using the \textit{Australia Telescope Compact Array (ATCA) \ion{H}{1} Galactic Center  Survey} \citep[see][for a recent update]{Teodoro2018}. They found a total of 86 clouds with LSR velocity $-209$ \kms\ $\leq v_{LSR} \leq +200$ \kms, and proposed that the clouds could be associated with an outflow from a star-forming region near the GC , and that the clouds are entrained material traveling at $\approx$ 200 \kms\ in a Galactic wind

Interestingly, the maximum absolute LSR velocity identified in their sample is 209 \kms. Comparing to our plot of maximal LSR velocity vs latitude (Figure~\ref{fig:velvglat}), we see that the \ion{H}{1} clouds do not follow the trend. In the range $20\degr \leq |b| \leq 30\degr$, the LSR velocity trend passes above the 200 \kms\ threshold. Thus, there is a substantial discrepancy between the LSR velocities of the HVCs seen in UV absorption and the clouds seen in \ion{H}{1}, and it appears that the HVCs in UV absorption are quantitatively distinct from the clouds in \ion{H}{1}. This may be related to the UV HVCs being seen at higher latitude, further from the plane. Furthermore, \citet{McClure-Griffiths2013} found that clouds at northern Galactic latitudes have predominantly positive velocities, while the ones in the south have predominantly negative velocities. As seen in Table~\ref{tab:allskytab},
we do not see any statistical difference between the number of positive and negative HVCs in the Southern Bubble, although B17 did detect more positive-velocity HVCs than negative-velocity HVCs in the Northern Bubble.

Another relevant \hi\ HVC is the Galactic Center Negative (GCN) complex. \citet{Winkel2011} present high-resolution \hi\ maps of this HVC complex, which forms a scattered population of $>$200 cloudlets covering $l$=0 to 50\degr\ and $b$=$-$40 to 0\degr\ with LSR velocities between $-$360 and $-$160~\kms\ (see their Figure 1). This region partially overlaps with the Southern Fermi Bubble, both spatially and kinematically. \citet{Winkel2011} conclude the Complex GCN consists of multiple sub-populations. It is possible that some of the gas in Complex GCN is related to the southern Fermi Bubble. However, none of the COS sightlines in our survey show 21\,cm emission in the GASS survey (Section \ref{sec:h1limits}), so Complex GCN is not a contaminant for our measurements.

\section{Conclusions}
\label{sec:conclusion}
We have presented \hst/COS FUV spectra of 17 AGN sightlines in the southern GC  region, six passing through the Southern Fermi Bubble (of which three pass near its boundary), and 11 passing outside (but nearby on the sky). The goal of this study was to search for absorbing gas associated with the GC  outflow and the Fermi Bubbles, and more broadly to identify the UV-absorption signature of the Fermi Bubble. We identified and removed foreground components originating in the Galactic disk and background components originating in the Magellanic Stream. Our principal observational findings are as follows:

\begin{enumerate}

\item Four of the six sightlines passing inside the Bubble show HVC detections,
whereas 
six of the eleven sightlines outside the Bubble show HVCs. The GC HVCs are detected in multiple UV species, including both low ions (\ion{C}{2}, \ion{Si}{2}, \ion{Al}{2}, \ion{Si}{3}) and high ions (\ion{C}{4}, \ion{Si}{4}).

\item An adjusted-$\chi^2$ test including Yates's correction for continuity shows no statistically significant enhancement in HVC covering fraction within the Southern Fermi Bubble. However, we did find a statistically significant enhancement when we combined the Southern Fermi Bubble sample with the Northern Fermi Bubble sample, at p-value 0.004 for blueshifted HVCs, and at p-value 0.02 when combining negative- and positive-velocity HVCs.
Because our sample size was $\approx$2.5 times smaller than that of B17, we require more data to make a more robust claim about the enhancement of HVCs within the Southern Bubble.

\item For sightlines within the Fermi Bubble (up to $|b|$=55\degr), we observe a decrease in the maximal absolute LSR velocity of the HVCs (i.e. the most positive or most negative velocity component) as a function of galactic latitude, from 280 \kms\ down to 120 \kms. This trend is seen in both the northern and the southern Bubbles, and for both blueshifted and redshifted components. This is expected if the GC  HVCs are tracing the nuclear outflow, since the gas should have higher velocity near the central engine and decelerated as it rises into the halo.

\item We measured the column-density ratios \ion{Si}{3}/\ion{Si}{2}, \ion{Si}{4}/\ion{Si}{3} and \ion{Si}{4}/\ion{Si}{2} in the GC  HVCs and compared them with 
latitude, to look for changes in ionization conditions with location in the outflow. We see 
no evidence for any significant correlations, although weak correlations may be present between \ion{Si}{3}/\ion{Si}{2} and $|b|$ and between \ion{Si}{4}/\ion{Si}{2} and $|b|$.
We require more data in the lower galactic latitude regime to further explore these trends and to confirm their existence. 
\end{enumerate}

Combined with the results from the detailed sight line analysis presented by \citet{Savage2017}, our results indicate the Southern Fermi Bubble is a dynamic, multi-phase environment giving rise to complex UV absorption-line profiles. Our study indicates that cool ($T\sim10^4$\,K) low-ionization clouds are able to survive in the hot Fermi Bubble plasma, perhaps as shredded pieces of interstellar material picked up and driven out by the hot nuclear wind. It is notable that despite the complex small-scale structure in the gas, the bulk properties of the clouds show discernible trends with latitude.




\acknowledgements
Support for programs 12936 and 13448 was provided by NASA through grants from the Space Telescope Science Institute, which is operated by the Association of Universities for Research in Astronomy, Inc., under NASA contract NAS 5-26555. This project was supported in part by the NSF REU grant AST-1358980 and by the Nantucket Maria Mitchell Association. Tanveer Karim would like to thank Eric G. Blackman for being his senior thesis internal adviser at the University of Rochester. Joss Bland-Hawthorn acknowledges an ARC Laureate Fellowship
and a Miller Professorship at UC Berkeley.\\

\facility{HST (COS), SALT (HRS), Parkes}\\
\software{Astropy \citep{astropy13}, Numpy \citep{Walt2011}, VPFIT \citep{VPFIT2014}, RDGEN \citep{RDGEN2014}}

\bibliographystyle{apj} 
\bibliography{apj-jour.bib,biblio.bib}

\appendix

\section{\hst/COS profiles for the Full Sample}
\label{sec:voigtprofiles} 
\begin{figure}
\label{fig:stackplots}
\plotone{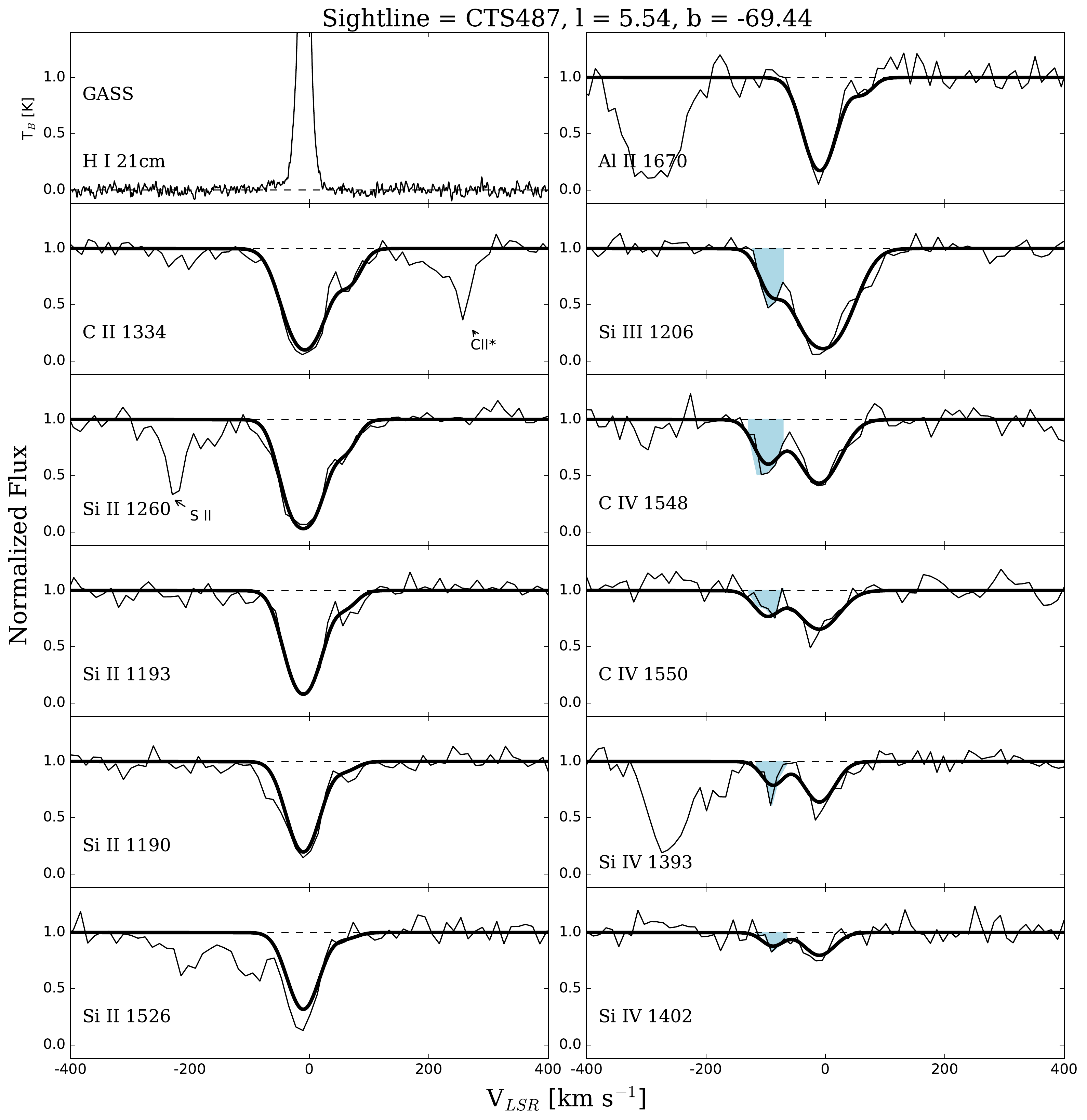}
\caption{\hst/COS absorption-line spectra for each direction in the sample (black lines), with VPFIT models overplotted in bold black. Directions without G160M spectra (G130M only) have reduced wavelength coverage, explaining the missing panels in some directions. The y-axis is the normalized flux and the x-axis is the LSR velocity in units of \kms. A 21\,cm \ion{H}{1} panel is included in the top-left panel. The complete figure set (15 images) is available in the online journal.}
\end{figure}

This appendix presents (Figure~\ref{fig:stackplots}) the \hst/COS spectra with the Voigt profile fits overplotted for all sightlines in the sample except the two directions (RBS 1768 and RBS 2000) already shown in Section~\ref{sec:results}. The parameters describing the component fits are given in Table~\ref{tab:vpfitresults}.

\figsetstart
\figsettitle{HST/COS Profiles for the Sightlines}

\figsetgrpstart
\figsetgrpnum{6.1}
\figsetgrptitle{CTS 487
}
\figsetplot{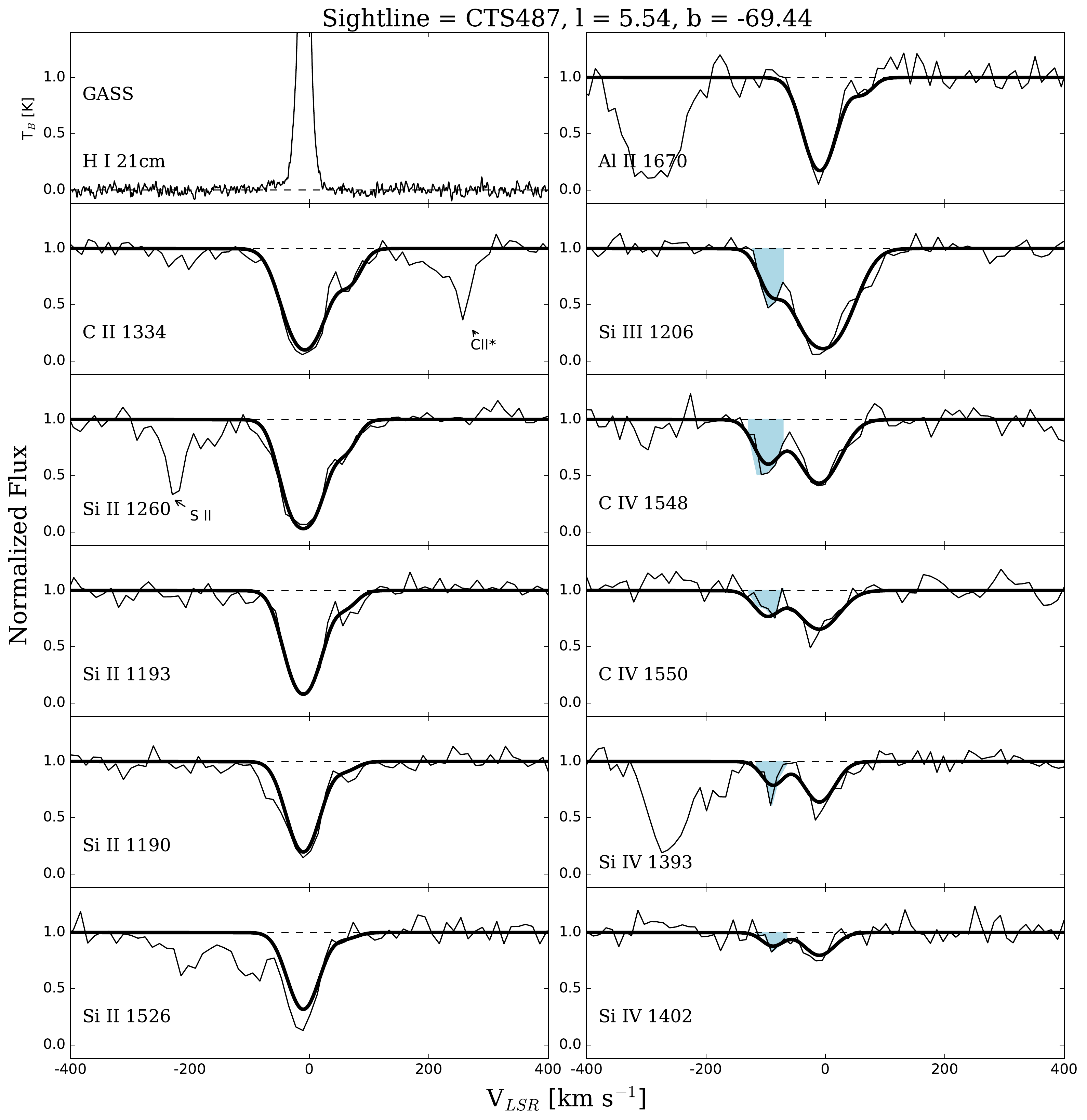}
\figsetgrpnote{\hst/COS absorption-line spectra for each direction in the sample (black lines), with VPFIT models overplotted in bold black. Directions without G160M spectra (G130M only) have reduced wavelength coverage, explaining the missing panels in some directions. The y-axis is the normalized flux and the x-axis is the LSR velocity in units of \kms. A 21\,cm \ion{H}{1} panel is included in the top-left panel.}
\figsetgrpend

\figsetgrpstart
\figsetgrpnum{6.2}
\figsetgrptitle{ESO 141-G55
}
\figsetplot{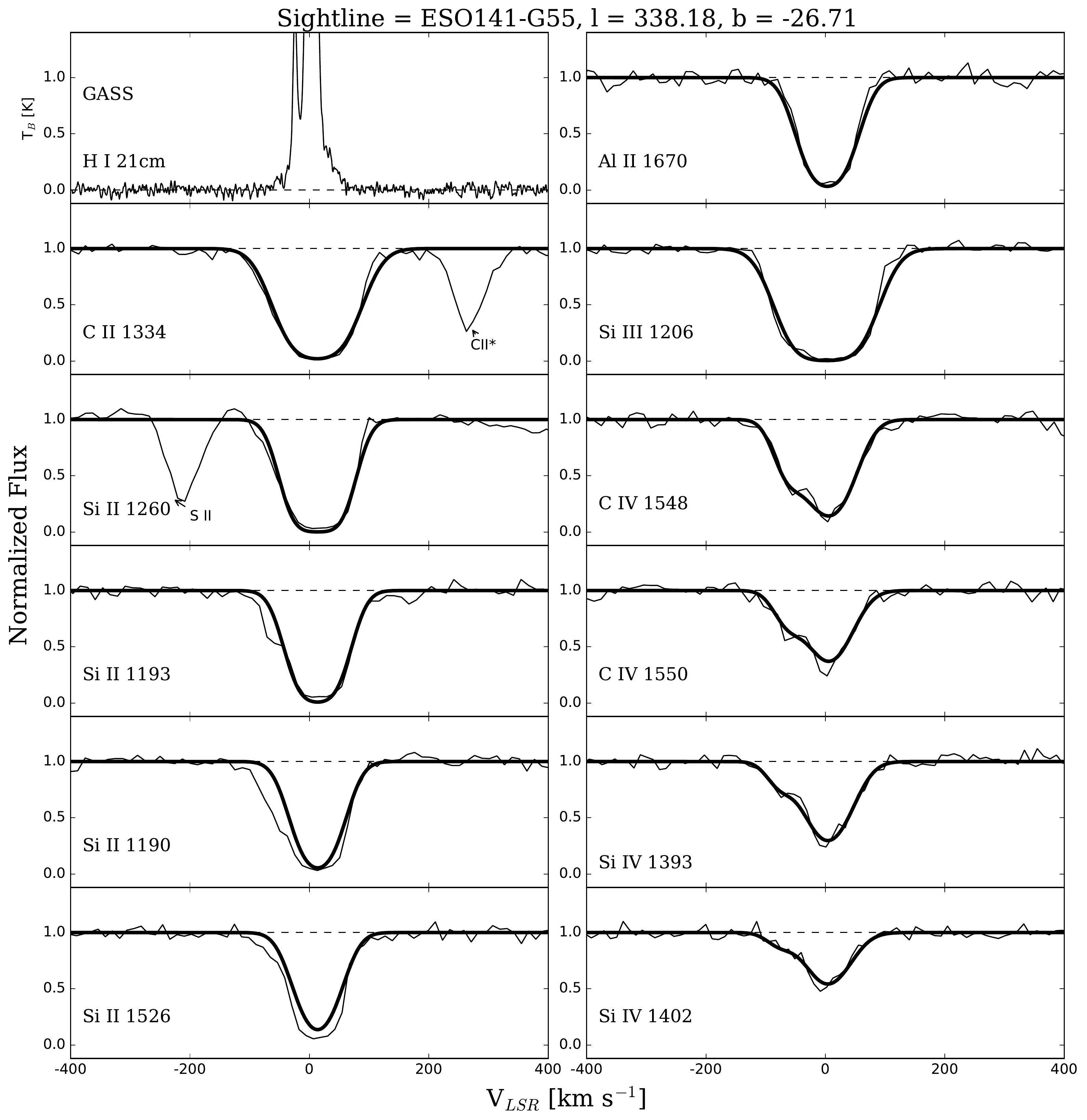}
\figsetgrpnote{\hst/COS absorption-line spectra for each direction in the sample (black lines), with VPFIT models overplotted in bold black. Directions without G160M spectra (G130M only) have reduced wavelength coverage, explaining the missing panels in some directions. The y-axis is the normalized flux and the x-axis is the LSR velocity in units of \kms. A 21\,cm \ion{H}{1} panel is included in the top-left panel.}
\figsetgrpend

\figsetgrpstart
\figsetgrpnum{6.3}
\figsetgrptitle{ESO 462-G09
}
\figsetplot{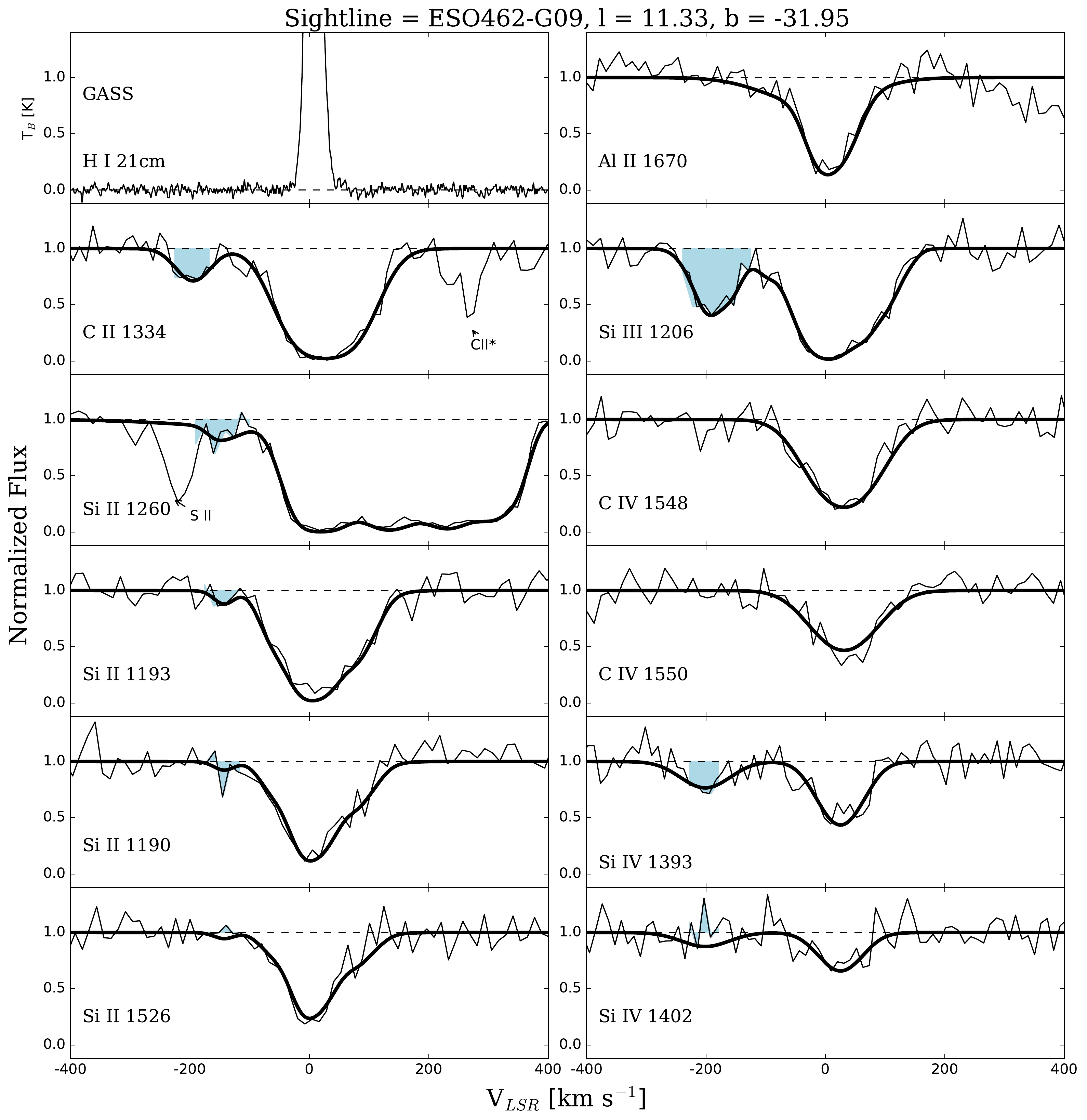}
\figsetgrpnote{\hst/COS absorption-line spectra for each direction in the sample (black lines), with VPFIT models overplotted in bold black. Directions without G160M spectra (G130M only) have reduced wavelength coverage, explaining the missing panels in some directions. The y-axis is the normalized flux and the x-axis is the LSR velocity in units of \kms. A 21\,cm \ion{H}{1} panel is included in the top-left panel.}
\figsetgrpend

\figsetgrpstart
\figsetgrpnum{6.4}
\figsetgrptitle{HE 2258-5524
}
\figsetplot{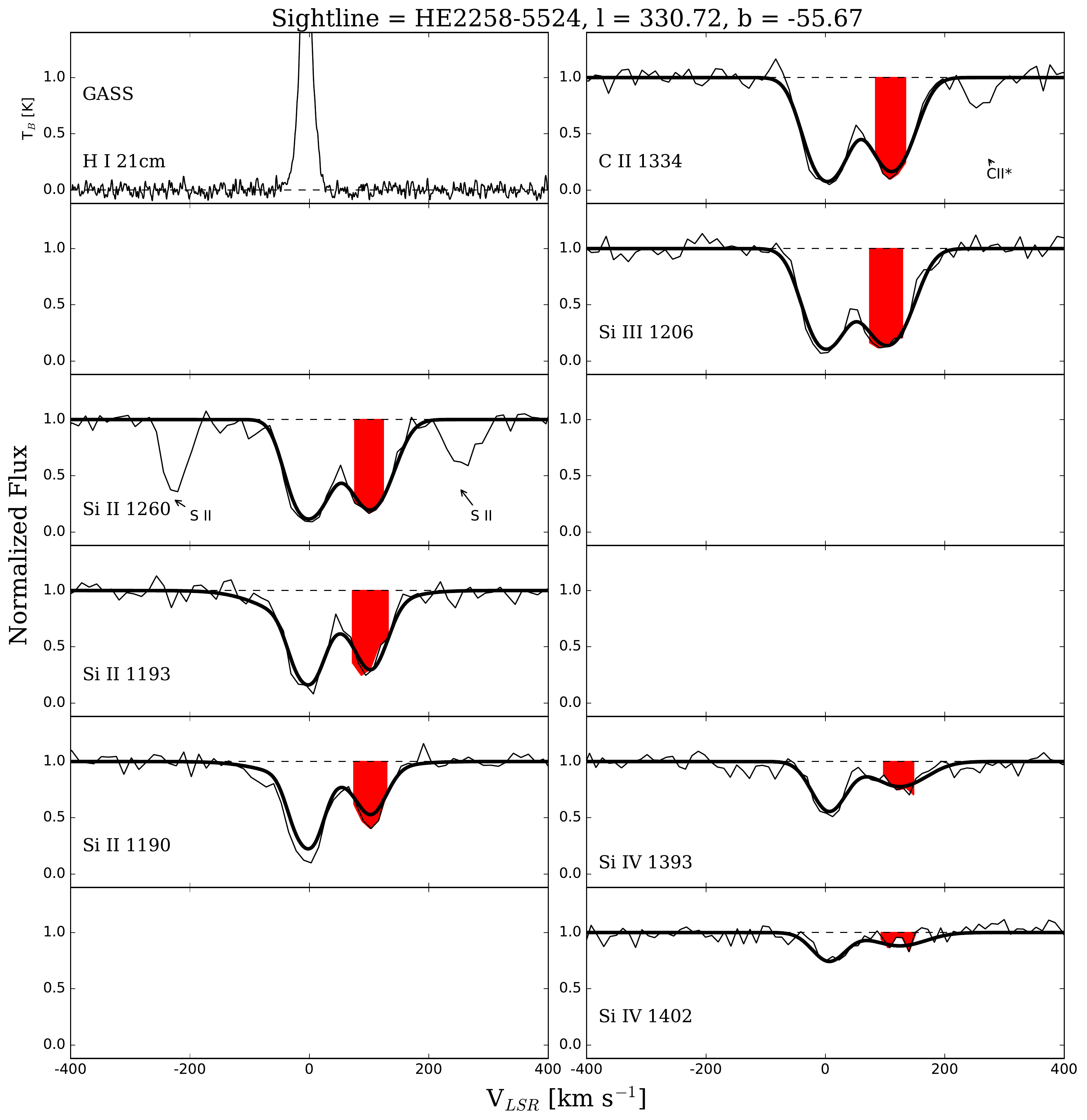}
\figsetgrpnote{\hst/COS absorption-line spectra for each direction in the sample (black lines), with VPFIT models overplotted in bold black. Directions without G160M spectra (G130M only) have reduced wavelength coverage, explaining the missing panels in some directions. The y-axis is the normalized flux and the x-axis is the LSR velocity in units of \kms. A 21\,cm \ion{H}{1} panel is included in the top-left panel.}
\figsetgrpend

\figsetgrpstart
\figsetgrpnum{6.5}
\figsetgrptitle{HE 2259-5524
}
\figsetplot{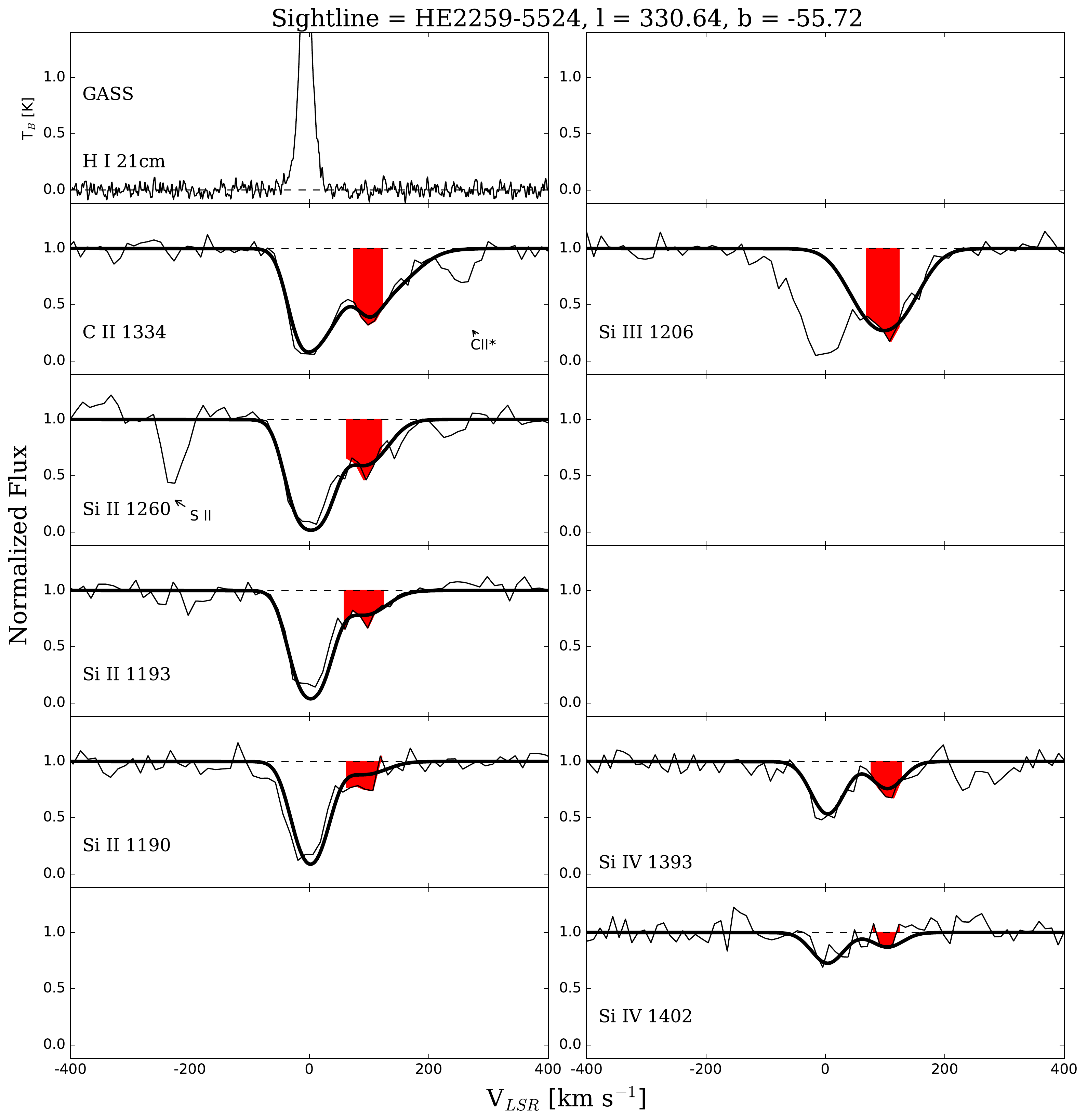}
\figsetgrpnote{\hst/COS absorption-line spectra for each direction in the sample (black lines), with VPFIT models overplotted in bold black. Directions without G160M spectra (G130M only) have reduced wavelength coverage, explaining the missing panels in some directions. The y-axis is the normalized flux and the x-axis is the LSR velocity in units of \kms. A 21\,cm \ion{H}{1} panel is included in the top-left panel.}
\figsetgrpend

\figsetgrpstart
\figsetgrpnum{6.6}
\figsetgrptitle{HE 2332-3556
}
\figsetplot{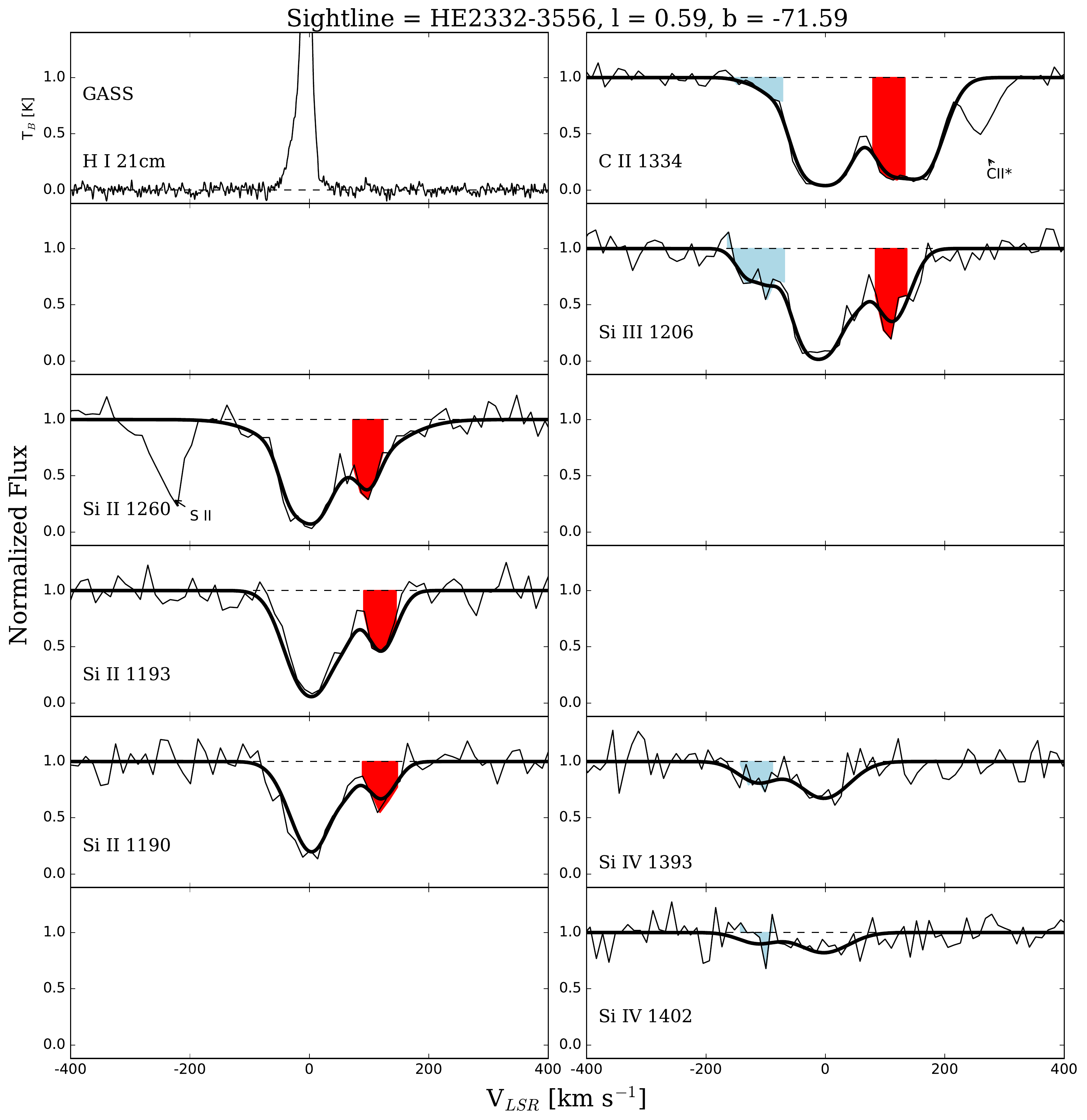}
\figsetgrpnote{\hst/COS absorption-line spectra for each direction in the sample (black lines), with VPFIT models overplotted in bold black. Directions without G160M spectra (G130M only) have reduced wavelength coverage, explaining the missing panels in some directions. The y-axis is the normalized flux and the x-axis is the LSR velocity in units of \kms. A 21\,cm \ion{H}{1} panel is included in the top-left panel.}
\figsetgrpend

\figsetgrpstart
\figsetgrpnum{6.7}
\figsetgrptitle{IRAS F21325-6
}
\figsetplot{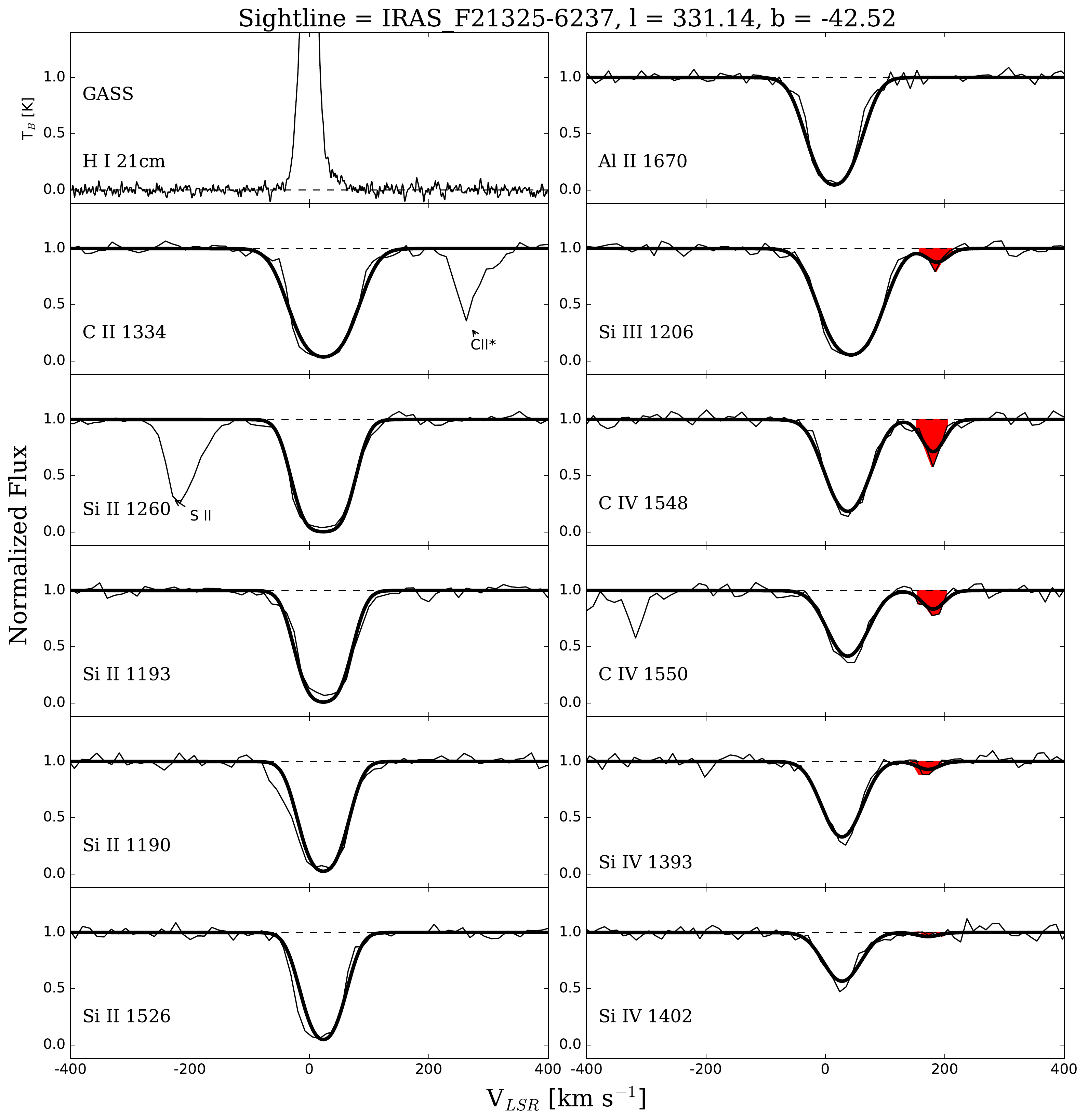}
\figsetgrpnote{\hst/COS absorption-line spectra for each direction in the sample (black lines), with VPFIT models overplotted in bold black. Directions without G160M spectra (G130M only) have reduced wavelength coverage, explaining the missing panels in some directions. The y-axis is the normalized flux and the x-axis is the LSR velocity in units of \kms. A 21\,cm \ion{H}{1} panel is included in the top-left panel.}
\figsetgrpend

\figsetgrpstart
\figsetgrpnum{6.8}
\figsetgrptitle{PKS 2005-489
}
\figsetplot{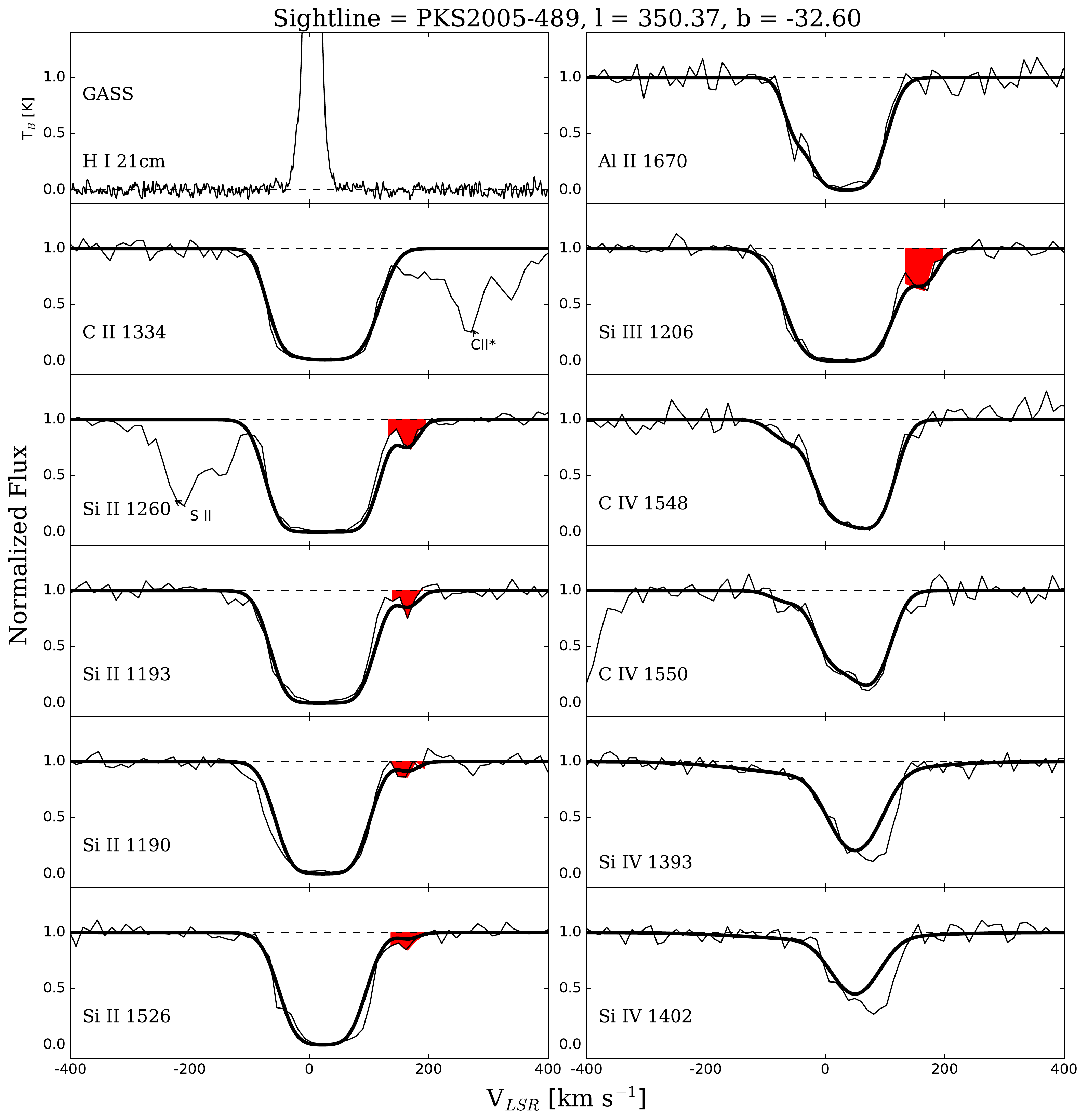}
\figsetgrpnote{\hst/COS absorption-line spectra for each direction in the sample (black lines), with VPFIT models overplotted in bold black. Directions without G160M spectra (G130M only) have reduced wavelength coverage, explaining the missing panels in some directions. The y-axis is the normalized flux and the x-axis is the LSR velocity in units of \kms. A 21\,cm \ion{H}{1} panel is included in the top-left panel.}
\figsetgrpend

\figsetgrpstart
\figsetgrpnum{6.9}
\figsetgrptitle{PKS 2155-304
}
\figsetplot{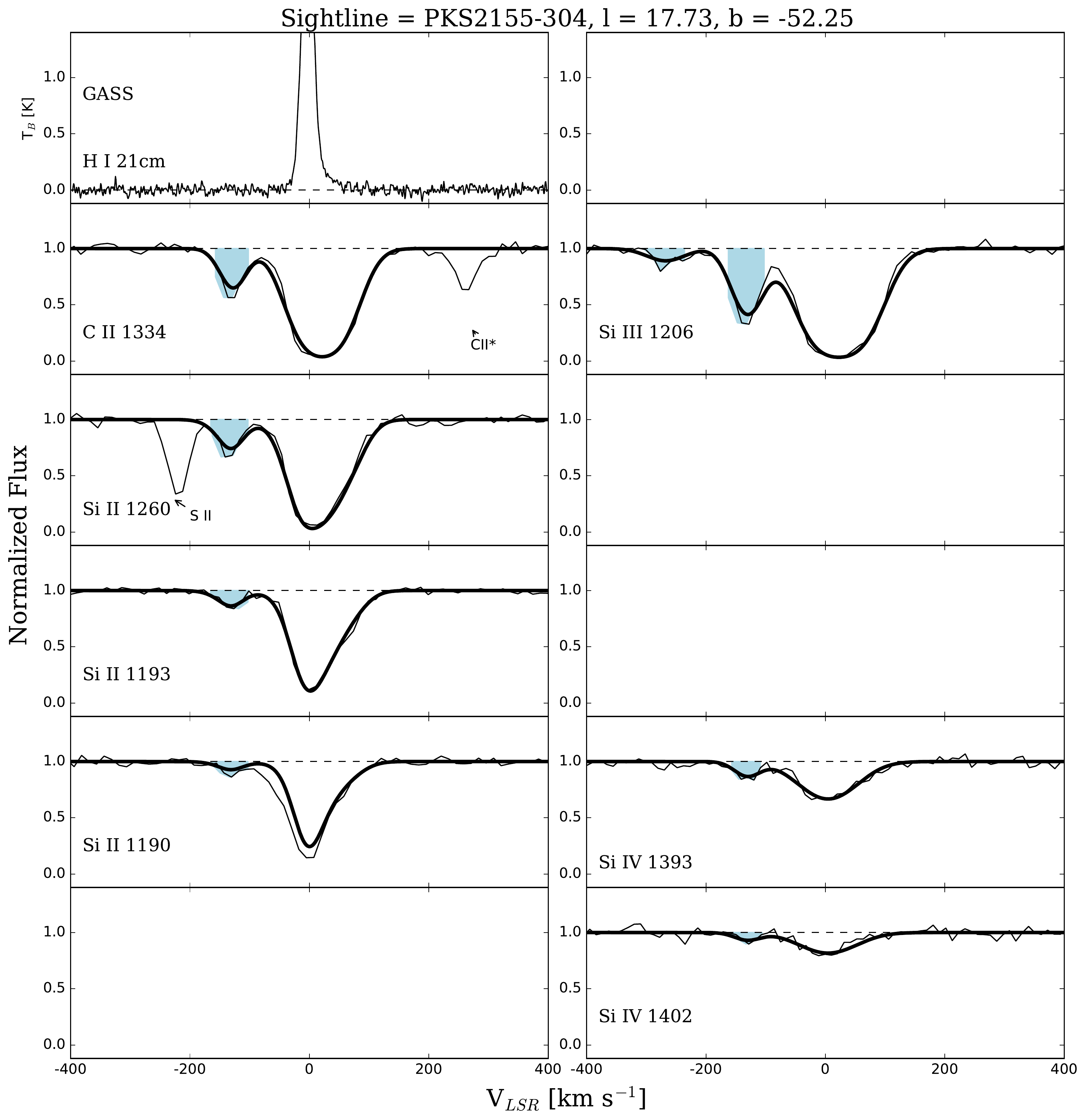}
\figsetgrpnote{\hst/COS absorption-line spectra for each direction in the sample (black lines), with VPFIT models overplotted in bold black. Directions without G160M spectra (G130M only) have reduced wavelength coverage, explaining the missing panels in some directions. The y-axis is the normalized flux and the x-axis is the LSR velocity in units of \kms. A 21\,cm \ion{H}{1} panel is included in the top-left panel.}
\figsetgrpend

\figsetgrpstart
\figsetgrpnum{6.10}
\figsetgrptitle{RBS 1666
}
\figsetplot{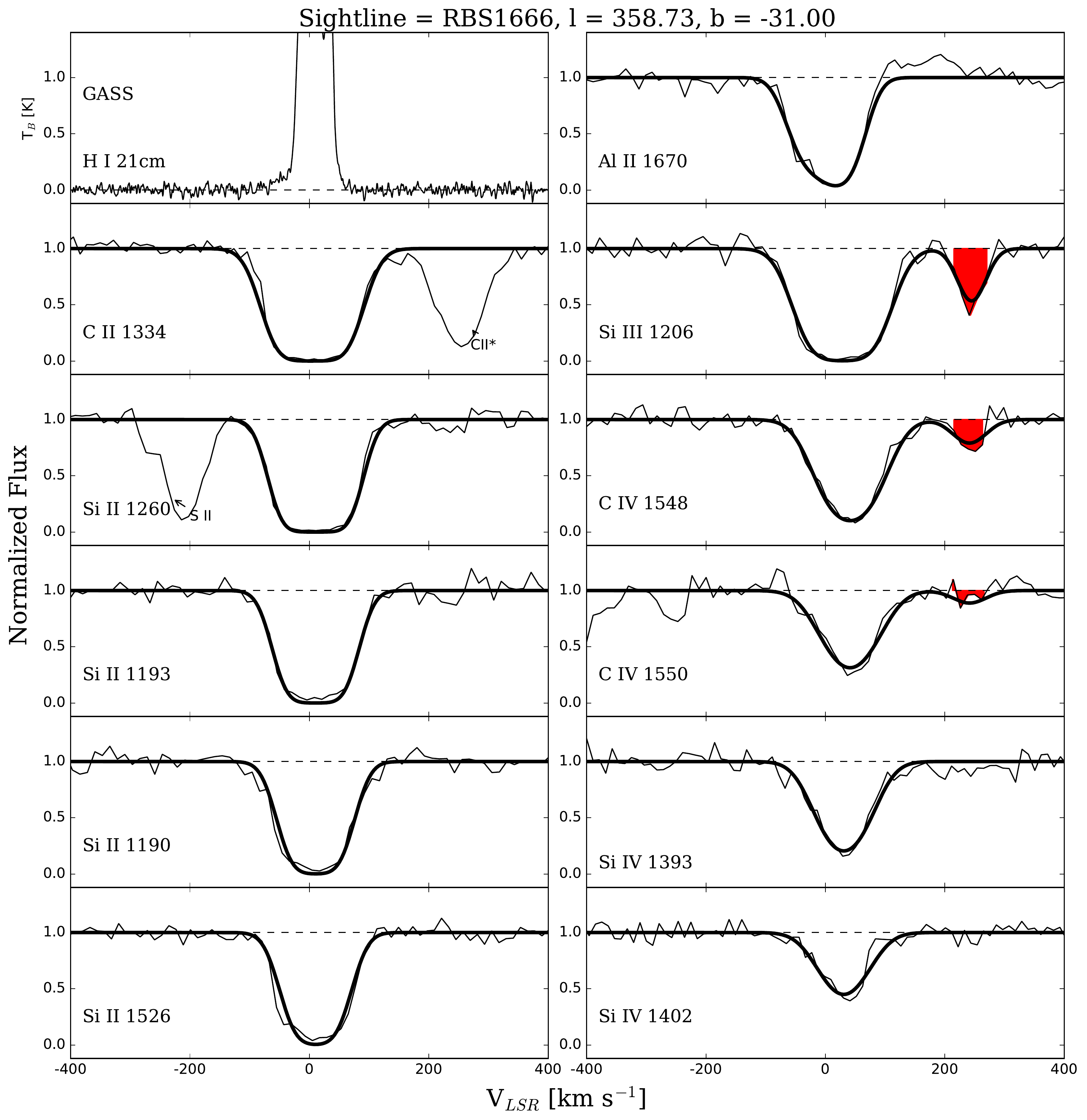}
\figsetgrpnote{\hst/COS absorption-line spectra for each direction in the sample (black lines), with VPFIT models overplotted in bold black. Directions without G160M spectra (G130M only) have reduced wavelength coverage, explaining the missing panels in some directions. The y-axis is the normalized flux and the x-axis is the LSR velocity in units of \kms. A 21\,cm \ion{H}{1} panel is included in the top-left panel.}
\figsetgrpend

\figsetgrpstart
\figsetgrpnum{6.11}
\figsetgrptitle{RBS 1795
}
\figsetplot{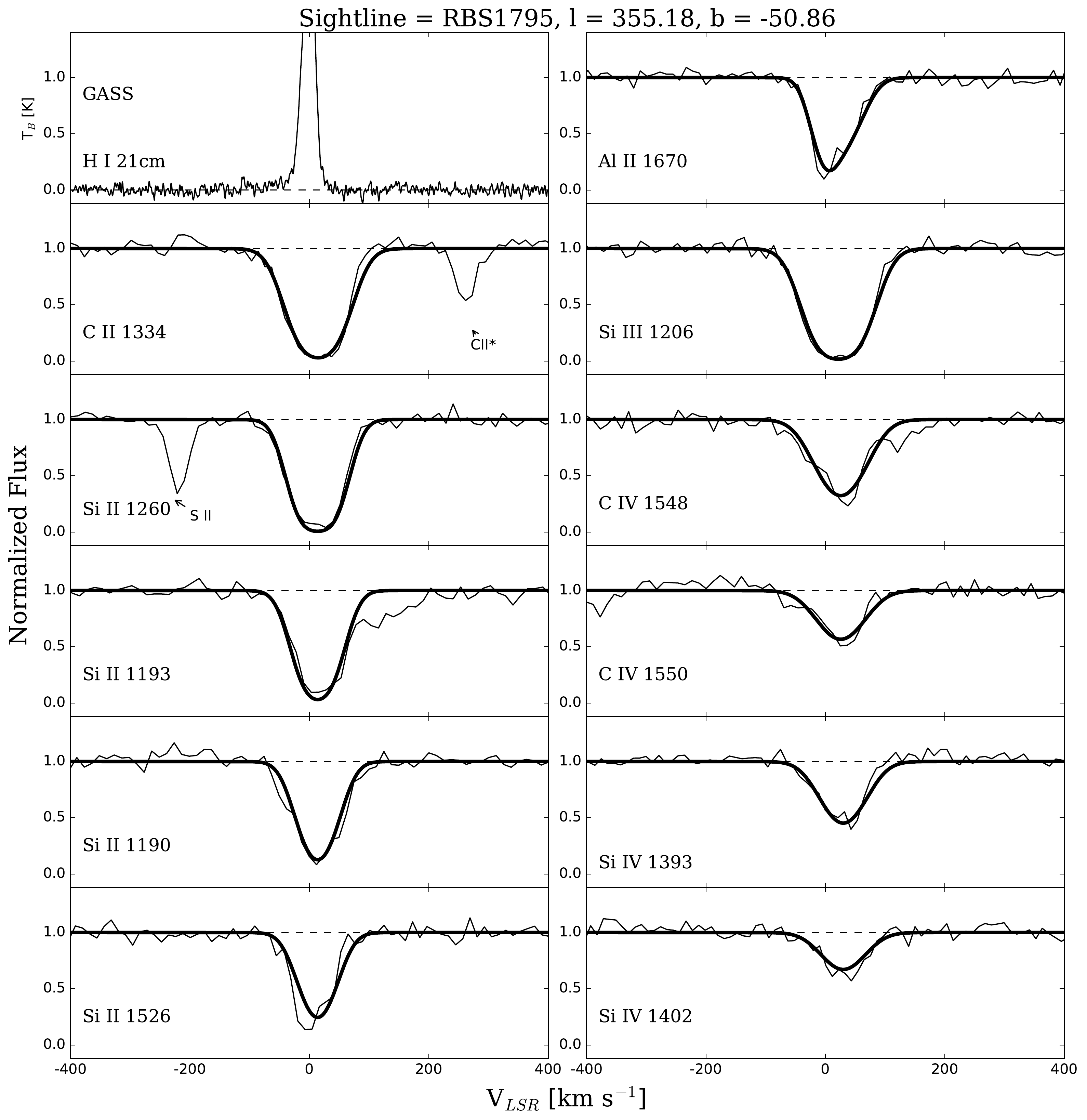}
\figsetgrpnote{\hst/COS absorption-line spectra for each direction in the sample (black lines), with VPFIT models overplotted in bold black. Directions without G160M spectra (G130M only) have reduced wavelength coverage, explaining the missing panels in some directions. The y-axis is the normalized flux and the x-axis is the LSR velocity in units of \kms. A 21\,cm \ion{H}{1} panel is included in the top-left panel.}
\figsetgrpend

\figsetgrpstart
\figsetgrpnum{6.12}
\figsetgrptitle{RBS 1892
}
\figsetplot{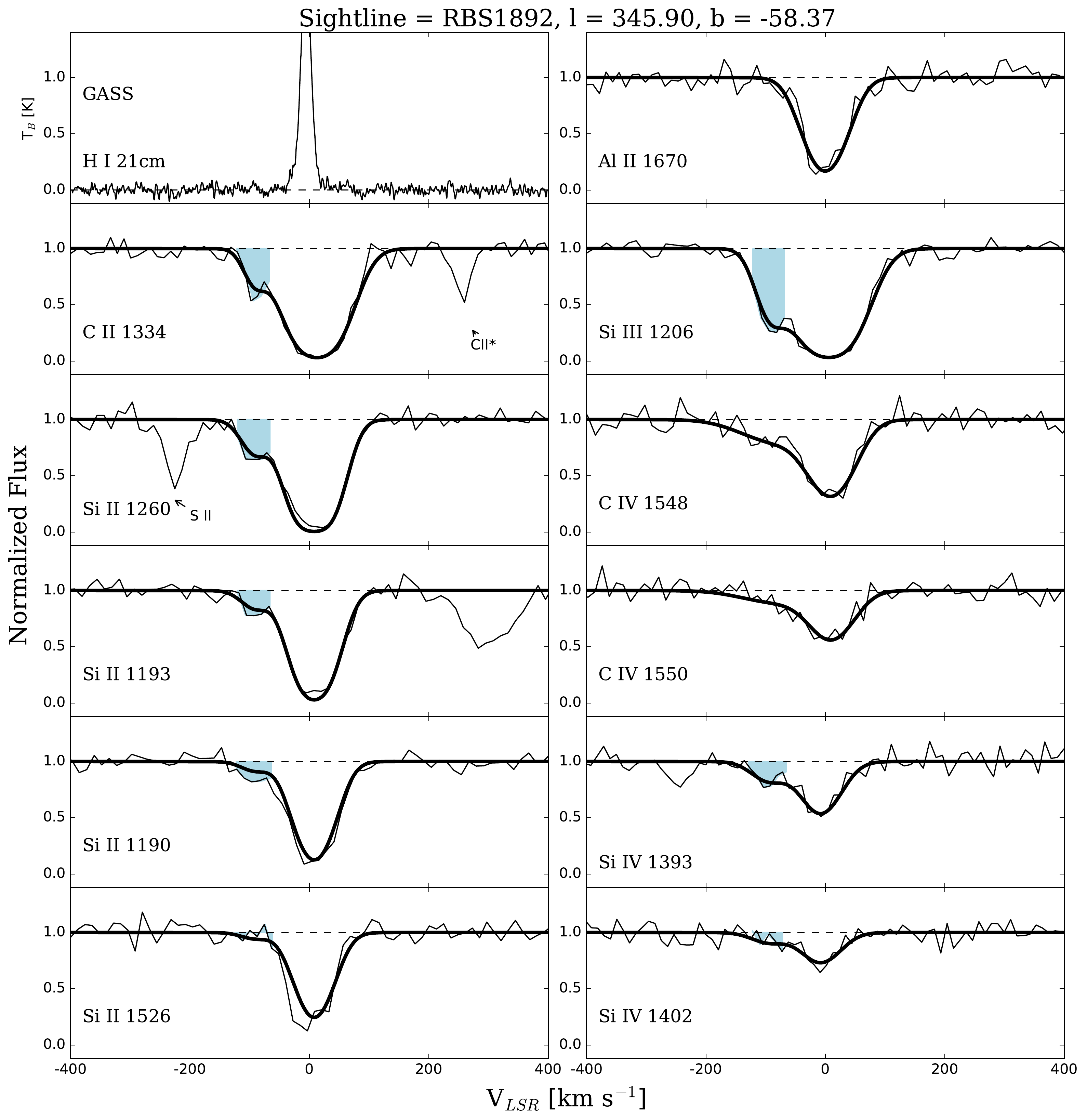}
\figsetgrpnote{\hst/COS absorption-line spectra for each direction in the sample (black lines), with VPFIT models overplotted in bold black. Directions without G160M spectra (G130M only) have reduced wavelength coverage, explaining the missing panels in some directions. The y-axis is the normalized flux and the x-axis is the LSR velocity in units of \kms. A 21\,cm \ion{H}{1} panel is included in the top-left panel.}
\figsetgrpend

\figsetgrpstart
\figsetgrpnum{6.13}
\figsetgrptitle{RBS 1897
}
\figsetplot{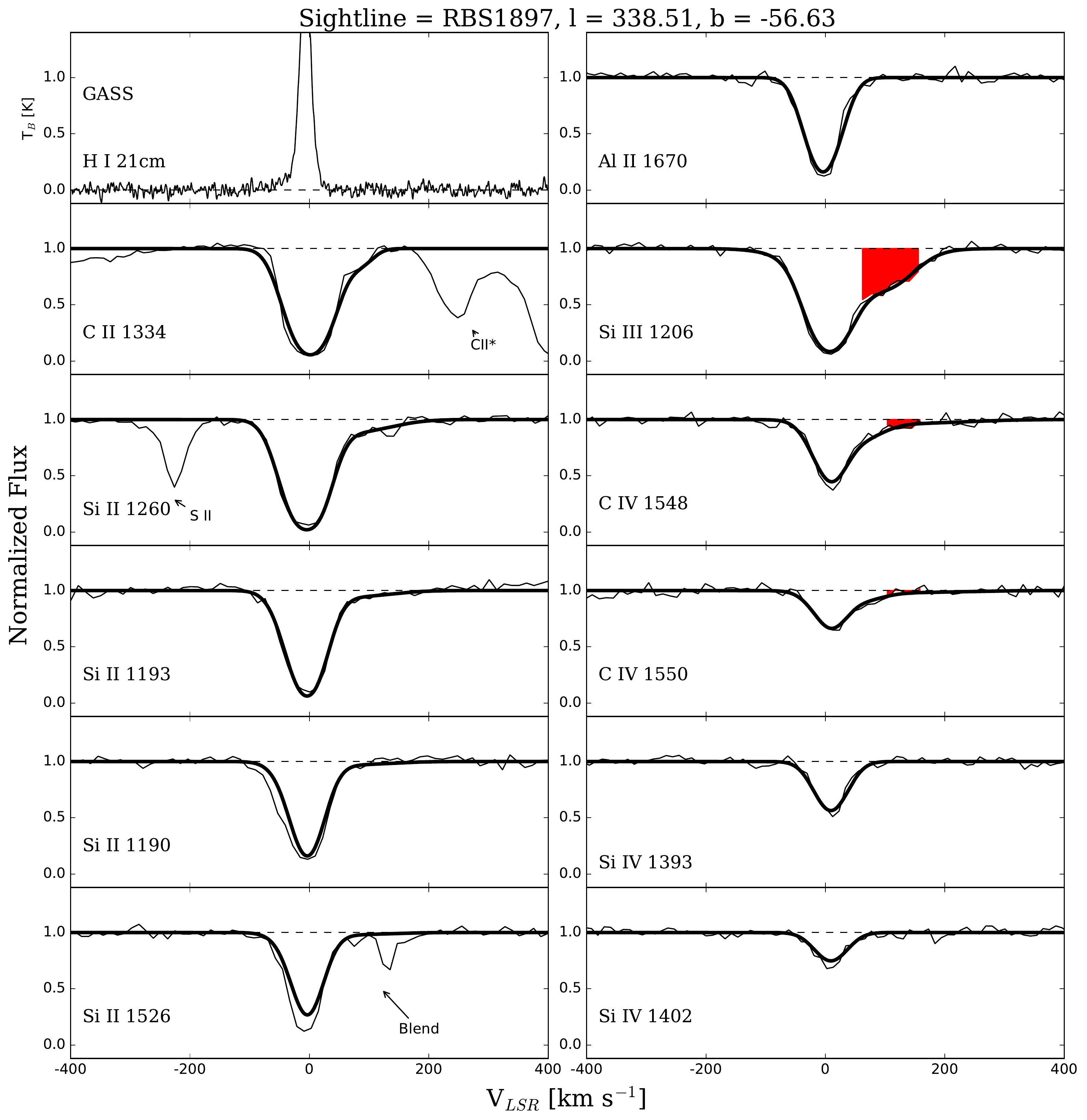}
\figsetgrpnote{\hst/COS absorption-line spectra for each direction in the sample (black lines), with VPFIT models overplotted in bold black. Directions without G160M spectra (G130M only) have reduced wavelength coverage, explaining the missing panels in some directions. The y-axis is the normalized flux and the x-axis is the LSR velocity in units of \kms. A 21\,cm \ion{H}{1} panel is included in the top-left panel.}
\figsetgrpend

\figsetgrpstart
\figsetgrpnum{6.14}
\figsetgrptitle{RBS 2023
}
\figsetplot{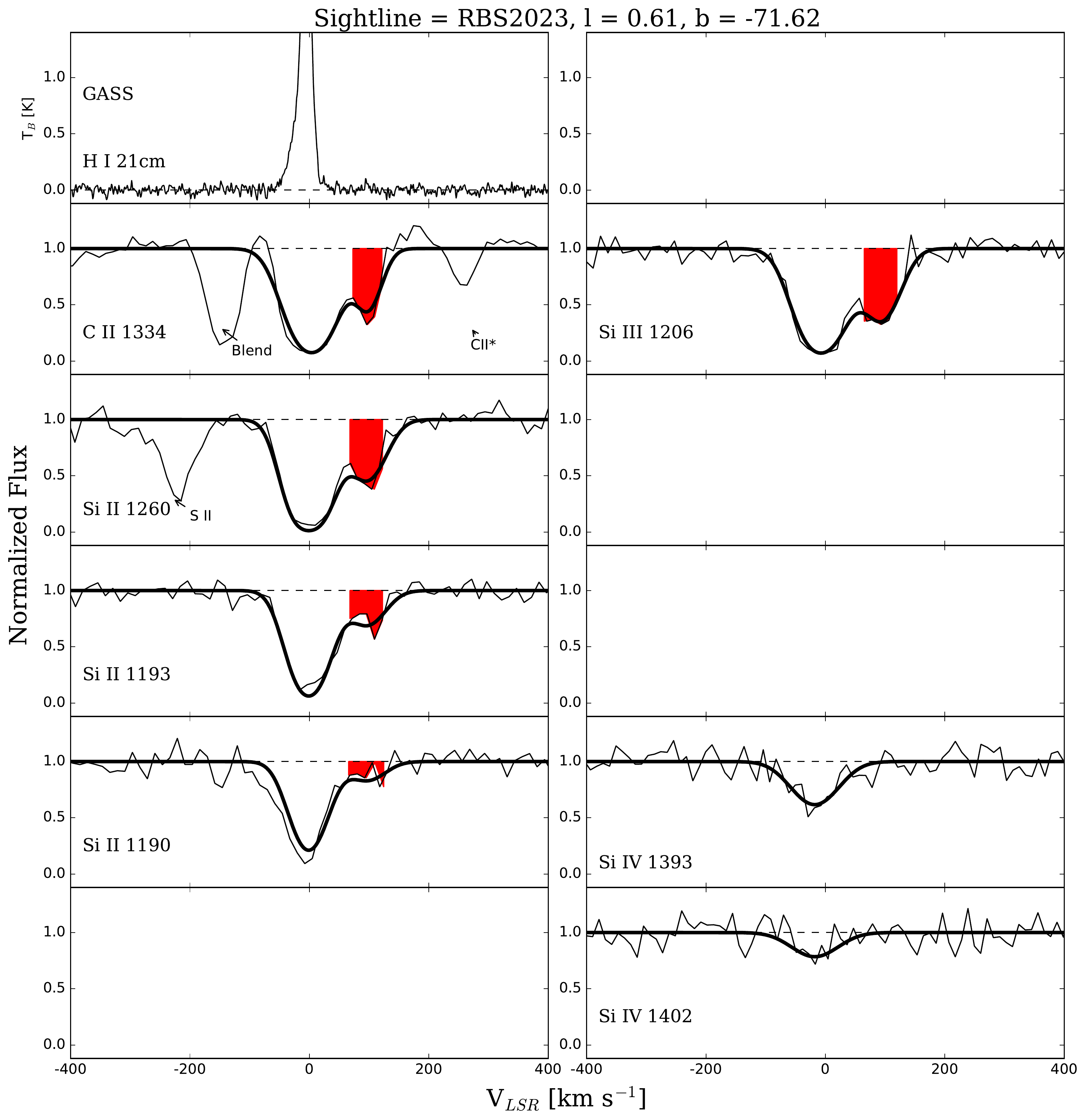}
\figsetgrpnote{\hst/COS absorption-line spectra for each direction in the sample (black lines), with VPFIT models overplotted in bold black. Directions without G160M spectra (G130M only) have reduced wavelength coverage, explaining the missing panels in some directions. The y-axis is the normalized flux and the x-axis is the LSR velocity in units of \kms. A 21\,cm \ion{H}{1} panel is included in the top-left panel.}
\figsetgrpend

\figsetgrpstart
\figsetgrpnum{6.15}
\figsetgrptitle{RBS 2070}
\figsetplot{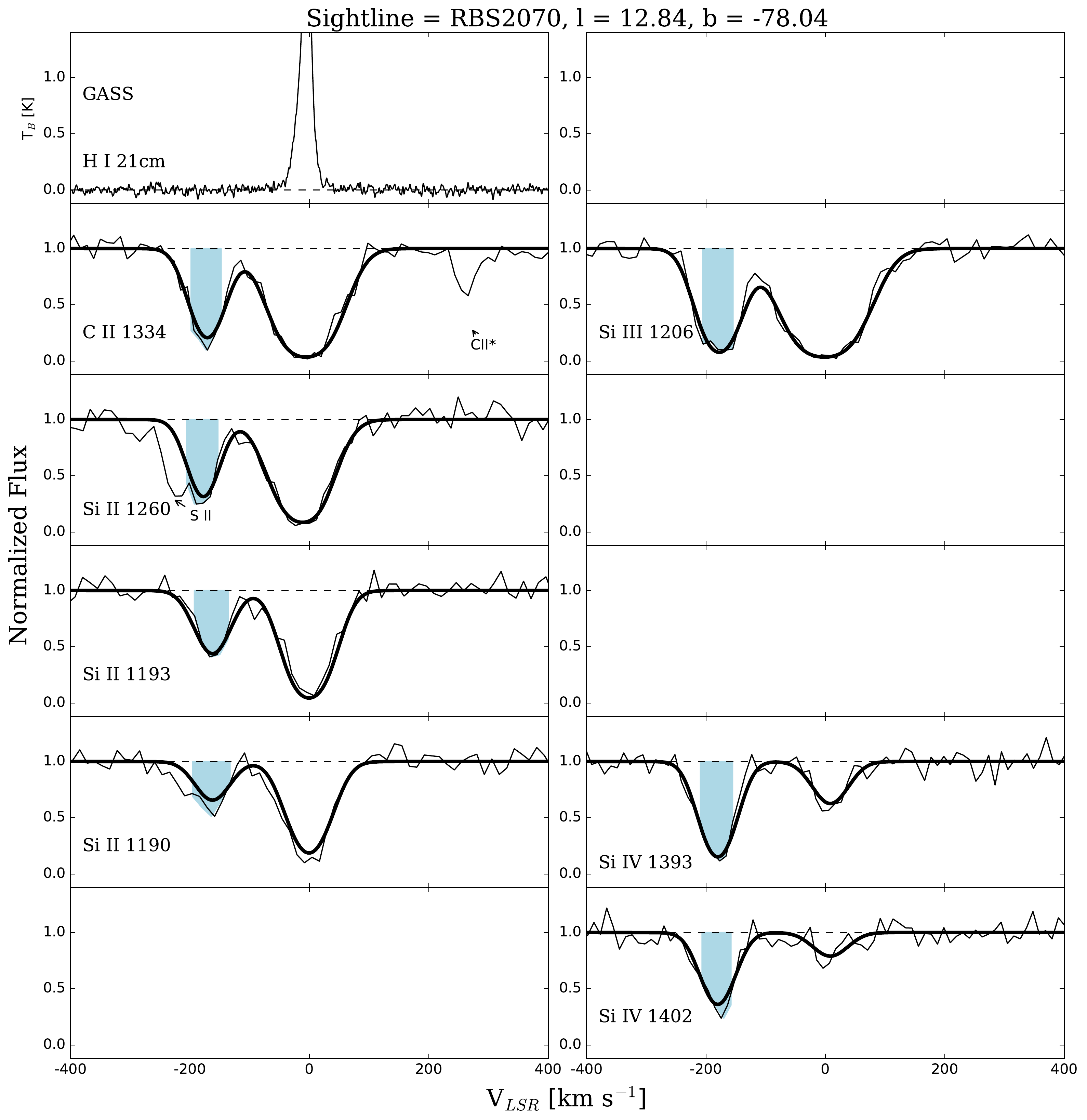}
\figsetgrpnote{\hst/COS absorption-line spectra for each direction in the sample (black lines), with VPFIT models overplotted in bold black. Directions without G160M spectra (G130M only) have reduced wavelength coverage, explaining the missing panels in some directions. The y-axis is the normalized flux and the x-axis is the LSR velocity in units of \kms. A 21\,cm \ion{H}{1} panel is included in the top-left panel.}
\figsetgrpend

\figsetend

\begin{deluxetable}{lllllrrrr}[h]
\tablecaption{HVC parameters derived from VPFIT models\label{tab:vpfitresults}\tm{a}}
\tabletypesize{\tiny}
\tablehead{\colhead{Sightline} & \colhead{$l$} & \colhead{$b$} & \colhead{Transition} & \colhead{$\Delta v$(Galactic)\tm{b}} & \colhead{$v_0$\tm{c}} & \colhead{$b$\tm{c}} & \colhead{log $N$\tm{c}} & \colhead{Flag\tm{d}}\\
\colhead{} & \colhead{($^{\circ}$)} & \colhead{($^{\circ}$)} & \colhead{} & \colhead{(\kms)} & \colhead{(\kms)} & \colhead{(\kms)} & \colhead{} }
\startdata
 & & & \ion{C}{4} 1548, 1550 & & $-$100 $\pm$ 2 & 18.1 $\pm$ 2.9 & 13.55 $\pm$ 0.04 & M\\
CTS 487 & 5.54 & $-$69.44 & \ion{Si}{3} 1206 & [0, 0.7] & $-$96 $\pm$ 1 & 10.6 $\pm$ 2.8 & 12.64 $\pm$ 0.05 & M\\
 &  &  & \ion{Si}{4} 1393, 1402 & & $-$90 $\pm$ 4 & 8.2 $\pm$ 8.0 & 12.81 $\pm$ 0.12 & M\\
\hline
& & &  \ion{Si}{4} 1393, 1402 & & $-$203 $\pm$ 8 & 50.7 $\pm$ 11.5 & 13.13 $\pm$ 0.08\\
& & & \ion{C}{2} 1334 & & $-$197 $\pm$ 6 & 31.6 $\pm$ 8.6 & 13.73 $\pm$ 0.09\\
ESO 462-G09 & 11.33 & $-$31.95 & \ion{Si}{3} 1206 & [0, 21.5] & $-$197 $\pm$ 14 & 14.2 $\pm$ 20.5 & 12.86 $\pm$ 1.05\\
& & & \ion{Si}{3} 1206 & & $-$159 $\pm$ 8 & 15.9 $\pm$ 9.8 & 12.71 $\pm$ 0.21\\
& & &  \ion{Si}{2} 1193, 1190 & & $-$145 $\pm$ 4 & 2.8 $\pm$ 5.0 & 12.75 $\pm$ 0.41\\
\hline
& & &  \ion{Si}{2} 1260& & 100 $\pm$ 1 & 37.0 $\pm$ 1.9 &  13.60 $\pm$ 0.02 & M\\
& & &  \ion{Si}{2} 1193, 1190 & & 102 $\pm$ 1 & 23.3 $\pm$ 2.4 &  13.62 $\pm$ 0.05 & M\\
HE2258-5524 & 330.72 & $-$55.67 & \ion{Si}{3} 1206 & [$-$10.2, 0] & 102 $\pm$ 1 & 40.7  $\pm$ 1.7 & 13.51  $\pm$ 0.02 & M\\
& & &  \ion{C}{2} 1334 & & 109 $\pm$ 1 & 36.4 $\pm$ 1.5 & 14.51 $\pm$ 0.02 & M\\
& & &  \ion{Si}{4} 1393, 1402 & & 123 $\pm$ 4 & 55.1 $\pm$ 6.4 & 13.14 $\pm$ 0.04 & M \\
\hline
& & &  \ion{Si}{2} 1260, 1193, 1190 & & 91 $\pm$ 3 & 42.1 $\pm$ 5.0 & 13.13 $\pm$ 0.04 & M\\
& & & \ion{Si}{3} 1206 & & 96 $\pm$ 3 & 55.9   $\pm$   3.9  & 13.43  $\pm$  0.02 & M\\
HE2259-5524 & 330.64 & -55.72 & \ion{C}{2} 1334 & [$-$10.2, 0] & 98 $\pm$ 5 & 15.5  $\pm$ 13.4 & 13.70 $\pm$ 0.65 & M \\
& & &  \ion{Si}{4} 1393, 1402 & & 102 $\pm$ 4 & 26.8 $\pm$     5.3 & 12.96 $\pm$   0.06 & M \\
\hline
& & & \ion{Si}{3} 1206 & & $-$133 $\pm$ 4 & 11.8 $\pm$ 7.4 & 12.44 $\pm$ 0.19\\
& & & \ion{C}{2} 1334 & & $-$129 $\pm$ 31 & 19.3 $\pm$ 42.0 & 12.70 $\pm$ 0.72\\
& & & \ion{Si}{4} 1393, 1402 & & $-$115 $\pm$ 9 & 37.0 $\pm$ 13.4 & 12.92 $\pm$ 0.13\\
& & & \ion{C}{2} 1334 & & $-$99 $\pm$ 5 & 5.9 $\pm$ 14.9 & 13.01 $\pm$ 0.28\\
HE 2332-3556 & 0.59 & $-$71.59 & \ion{Si}{3} 1206 & [0, 0.1] & $-$97 $\pm$ 2 & 2.8 $\pm$ 96.4 & 13.22 $\pm$ 0.10\\
& & & \ion{Si}{2} 1260 & & 98 $\pm$ 1 & 6.9 $\pm$ 5.0 & 13.16 $\pm$ 0.50\\
& & & \ion{C}{2} 1334 & & 106 $\pm$ 9 & 7.0 $\pm$ 15.9 & 13.93 $\pm$ 0.33\\
& & & \ion{Si}{3} 1206 & & 110 $\pm$ 2 & 28.8 $\pm$ 5.0 & 13.14 $\pm$ 0.10\\
& & & \ion{Si}{2} 1193, 1190 & & 118 $\pm$ 2 & 23.4 $\pm$ 3.3 & 13.50 $\pm$ 0.04\\
\hline
& & & \ion{Si}{4} 1393, 1402 & & 170 $\pm$ 3 & 9.1 $\pm$ 7.1 & 12.25 $\pm$ 0.10\\
IRAS F21325-6237 & 331.14 & $-$42.52 & \ion{C}{4} 1548, 1550 & [$-$21.6, 0] & 178 $\pm$ 1 & 11.5 $\pm$ 1.2 & 13.37 $\pm$ 0.02\\
& & & \ion{Si}{3} 1206 & & 185 $\pm$ 8 & 7.9 $\pm$ 4.2 & 12.08 $\pm$ 0.06\\
\hline
PKS 2005-489 & 350.37 & $-$32.60 & \ion{Si}{2} 1260, 1193, 1190, 1526 & [$-$18.3, 0] & 164 $\pm$ 2 & 5.3 $\pm$ 3.4 & 12.71 $\pm$ 0.18\\
& & & \ion{Si}{3} 1206 & & 165 $\pm$ 2 & 15.0 $\pm$ 3.0 & 12.52 $\pm$ 0.05\\
\hline
& & & \ion{Si}{3} 1206 & & $-$270 $\pm$ 5 & 36.5 $\pm$ 6.9 & 12.23 $\pm$ 0.06\\
& & & \ion{Si}{2} 1260, 1193, 1190 & & $-$134 $\pm$ 4 & 28.8 $\pm$ 12.6 & 12.56 $\pm$ 0.19 & M\\
PKS 2155-304 & 17.73 & $-$52.25 & \ion{Si}{3} 1206 & [0, 8.6] & $-$133 $\pm$ 1 & 26.3 $\pm$ 0.9 & 13.04 $\pm$ 0.01 & M\\
& & & \ion{Si}{4} 1393, 1402 & & $-$133 $\pm$ 1 & 18.6 $\pm$ 2.7 & 12.59 $\pm$ 0.04 & M\\
& & & \ion{C}{2} 1334 & & $-$130 $\pm$ 1 & 19.0 $\pm$ 2.1 & 13.74 $\pm$ 0.03 & M\\
\hline
RBS 1666 & 358.73 & $-$31.00 & \ion{C}{4} 1548, 1550 & [$-$2.9, 0] & 239 $\pm$ 2 & 30.0 $\pm$ 3.6 & 13.32 $\pm$ 0.04\\
& & & \ion{Si}{3} 1206 & & 243 $\pm$ 1 & 20.9 $\pm$ 1.9 & 12.86 $\pm$ 0.03\\
\hline
& & & \ion{C}{2} 1334 & & $-$163 $\pm$ 13 & 26.5 $\pm$ 9.9 & 13.77 $\pm$ 0.28\\
& & & \ion{C}{4} 1548, 1550 &  & $-$161 $\pm$ 4 & 3.2 $\pm$ 9.7 & 12.91 $\pm$ 0.44\\
& & & \ion{Si}{4} 1393, 1402 &  & $-$154 $\pm$ 3 & 14.2 $\pm$ 3.7 & 12.76 $\pm$ 0.13\\
& & & \ion{Si}{2} 1193, 1190, 1526 &  & $-$146 $\pm$ 55 & 57.0 $\pm$ 34.4 & 13.05 $\pm$ 0.57\\ 
& & & \ion{Si}{2} 1260 &  & $-$142 $\pm$ 11 & 22.5 $\pm$ 9.9 & 13.04 $\pm$ 0.64\\ 
& & & \ion{Si}{3} 1206 &  & $-$142 $\pm$ 1 & 35.1 $\pm$ 1.0 & 13.52 $\pm$ 0.01\\
RBS 1768 & 4.51 & $-$48.46 & \ion{C}{2} 1334 & [0, 3.1] & $-$130 $\pm$ 5 & 20.0 $\pm$ 3.4 & 13.94 $\pm$ 0.19\\
& & & \ion{C}{4} 1548, 1550 &  & $-$126 $\pm$ 3 & 16.3 $\pm$ 5.1 & 13.41 $\pm$ 0.08\\
& & & \ion{Si}{4} 1393, 1402 &  & $-$119 $\pm$ 4 & 21.0 $\pm$ 4.9 & 12.92 $\pm$ 0.10\\
& & & \ion{Si}{2} 1260 &  & 148 $\pm$ 2 & 7.5 $\pm$ 5.2 & 12.31 $\pm$ 0.07\\ 
& & & \ion{Si}{3} 1206 &  & 148 $\pm$ 3 & 31.7 $\pm$ 4.8 & 12.53 $\pm$ 0.05\\
& & & \ion{C}{2} 1334 & & 154 $\pm$ 1 & 17.9 $\pm$ 2.4 & 13.53 $\pm$ 0.04\\
& & & \ion{Si}{2} 1193, 1190, 1526 &  & 157 $\pm$ 8 & 43.2 $\pm$ 13.3 & 12.87 $\pm$ 0.10\\ 
\hline
& & & \ion{Si}{4} 1393, 1402 &  & $-$97 $\pm$ 6 & 31.4 $\pm$ 9.4 & 12.84 $\pm$ 0.10\\
& & & \ion{Si}{3} 1206 & & $-$95 $\pm$ 1 & 21.8 $\pm$ 1.9 & 13.02 $\pm$ 0.03\\
RBS 1892 & 345.9 & $-$58.37 & \ion{C}{2} 1334 & [$-$4.5, 0] & $-$94 $\pm$ 2 & 10.2 $\pm$ 3.5 & 13.62 $\pm$ 0.06\\
& & & \ion{Si}{2} 1260, 1193, 1190, 1526 &  & $-$93 $\pm$ 2 & 21.4 $\pm$ 3.7 & 12.80 $\pm$ 0.05\\
\hline
RBS 1897 & 338.51 & $-$56.63 & \ion{Si}{3} 1206 & [-7.2, 0] & 109 $\pm$ 23 & 49.3 $\pm$ 26.6 & 12.58 $\pm$ 0.72 & M\\
& & & \ion{C}{4} 1548, 1550 & & 131 $\pm$ 226 & 126.6 $\pm$ 179.0 & 13.02 $\pm$ 0.99 & M\\
\enddata
\tn{a}{Each row in this table shows the output from VPFIT from component fitting to the listed transitions. Where multiple transitions from a given ion are listed, the data were fitted simultaneously. Only HVCs are listed ($|v_{\rm LSR}|>90$ \kms), not intermediate-velocity clouds (IVCs) or low-velocity clouds (LVCs); so directions in our sample without HVCs (ESO 141-G55 and RBS 1795) do not appear in this table. Some of the reported HVCs may have a more complex velocity structure than reported here due to a combination of broad and narrow feature, and/or convolution with blends. Thus, our modeled Voigt profile for these HVCs are an approximation and our modeled profiles may appear as underfits in Section~\ref{sec:voigtprofiles}.}
\tn{b}{Range of velocities allowed by model of differential Galactic rotation. For further discussion, see Section~\ref{sec:ism-blend}.}
\tn{c}{LSR Velocity centroid, Doppler $b$-parameter, and column density of component.}
\tn{d}{Flag M denotes Magellanic Stream absorption.}
\end{deluxetable}

\begin{deluxetable}{lllllrrrr}[h]
\tablecaption{HVC parameters derived from VPFIT models (Contd.)\tm{a}}
\tabletypesize{\tiny}
\tablehead{\colhead{Sightline} & \colhead{$l$} & \colhead{$b$} & \colhead{Transition} & \colhead{$\Delta v$(Galactic)\tm{b}} & \colhead{$v_0$\tm{c}} & \colhead{$b$\tm{c}} & \colhead{log $N$\tm{c}} & \colhead{Flag\tm{d}}\\
\colhead{} & \colhead{($^{\circ}$)} & \colhead{($^{\circ}$)} & \colhead{} & \colhead{(\kms)} & \colhead{(\kms)} & \colhead{(\kms)} & \colhead{} }
\startdata
& & & \ion{Si}{2} 1260 &  & $-$154 $\pm$ 8 & 30.0 $\pm$ 17.5 & 12.82 $\pm$ 0.18 & M\\ 
& & & \ion{C}{4} 1548, 1550 &  & $-$154 $\pm$ 1 & 18.3 $\pm$ 2.2 & 13.76 $\pm$ 0.04 & M\\
& & & \ion{Si}{3} 1206 &  & $-$150 $\pm$ 1 & 16.9 $\pm$ 1.3 & 13.04 $\pm$ 0.03 & M\\
& & & \ion{Si}{4} 1393, 1402 &  & $-$150 $\pm$ 1 & 15.8 $\pm$ 2.0 & 13.21 $\pm$ 0.03 & M\\
& & & \ion{C}{2} 1334 &  & $-$149 $\pm$ 1 & 15.0 $\pm$ 2.3 & 13.81 $\pm$ 0.04 & M\\
RBS 2000 & 350.20 & $-$67.58 & \ion{Si}{2} 1193, 1190, 1526 & [$-$1.5, 0] & $-$145 $\pm$ 6 & 14.0 $\pm$ 8.0 & 12.88 $\pm$ 0.18 & M\\ 
& & & \ion{C}{2} 1334 &  & 136 $\pm$ 1 & 16.6 $\pm$ 1.8 & 14.01 $\pm$ 0.03\\
& & & \ion{Si}{2} 1260 & & 136 $\pm$ 1 & 12.4 $\pm$ 1.4 & 13.08 $\pm$ 0.03\\ 
& & & \ion{Al}{2} 1670 & & 136 $\pm$ 2 & 7.0 $\pm$ 6.1 & 12.47 $\pm$ 0.25\\
& & & \ion{Si}{3} 1206 &  & 139 $\pm$ 1 & 20.5 $\pm$ 1.9 & 12.88 $\pm$ 0.03\\
& & & \ion{Si}{2} 1193, 1190, 1526 &  & 143 $\pm$ 4 & 14.5 $\pm$ 8.8 & 13.00 $\pm$ 0.39\\ 
\hline
& & & \ion{Si}{3} 1206 & & 92 $\pm$ 2 & 31.9 $\pm$ 2.9 & 13.15 $\pm$ 0.03\\
RBS 2023 & 0.61 & $-$71.62 & \ion{C}{2} 1334 & [0, 0.1] & 97 $\pm$ 1 & 17.5 $\pm$ 1.4 & 14.00 $\pm$ 0.02\\
& & & \ion{Si}{2} 1260, 1193, 1190 & & 95 $\pm$ 2 & 32.7 $\pm$ 3.1 & 13.23 $\pm$ 0.03\\
\hline
& & & \ion{Si}{4} 1393, 1402 & & $-$182 $\pm$ 1 & 28.6 $\pm$ 1.2 & 13.87 $\pm$ 0.02 & M\\
& & & \ion{Si}{2} 1260 &  & $-$180 $\pm$ 1 & 24.9 $\pm$ 1.9 & 13.37 $\pm$ 0.02 & M\\
RBS 2070 & 12.84 & $-$78.04 & \ion{Si}{3} 1206 & [0, 0.5] & $-$180 $\pm$ 1 & 33.9 $\pm$ 1.5 & 13.60 $\pm$ 0.03 & M\\
& & & \ion{C}{2} 1334 &  & $-$172 $\pm$ 1 & 29.2 $\pm$ 1.4 & 14.41 $\pm$ 0.02 & M\\
& & & \ion{Si}{2} 1193, 1190 &  & $-$164 $\pm$ 2 & 30.8 $\pm$ 2.1 & 13.57 $\pm$ 0.03 & M\\
\enddata
\tn{a}{Each row in this table shows the output from VPFIT from component fitting to the listed transitions. Where multiple transitions from a given ion are listed, the data were fitted simultaneously. Only HVCs are listed ($|v_{\rm LSR}|>90$ \kms), not intermediate-velocity clouds (IVCs) or low-velocity clouds (LVCs); so directions in our sample without HVCs (ESO 141-G55 and RBS 1795) do not appear in this table. Some of the reported HVCs may have a more complex velocity structure than reported here due to a combination of broad and narrow feature, and/or convolution with blends. Thus, our modeled Voigt profile for these HVCs are an approximation and our modeled profiles may appear as underfits in Section~\ref{sec:voigtprofiles}. }
\tn{b}{Range of velocities allowed by model of differential Galactic rotation. For further discussion, see Section~\ref{sec:ism-blend}.}
\tn{c}{LSR Velocity centroid, Doppler $b$-parameter, and column density of component.}
\tn{d}{Flag M denotes Magellanic Stream absorption.}
\end{deluxetable}

\clearpage
\section{SALT/HRS observations}
\label{sec:saltsec}
\begin{figure*}[!h]
    \epsscale{1.0}
    \plottwo{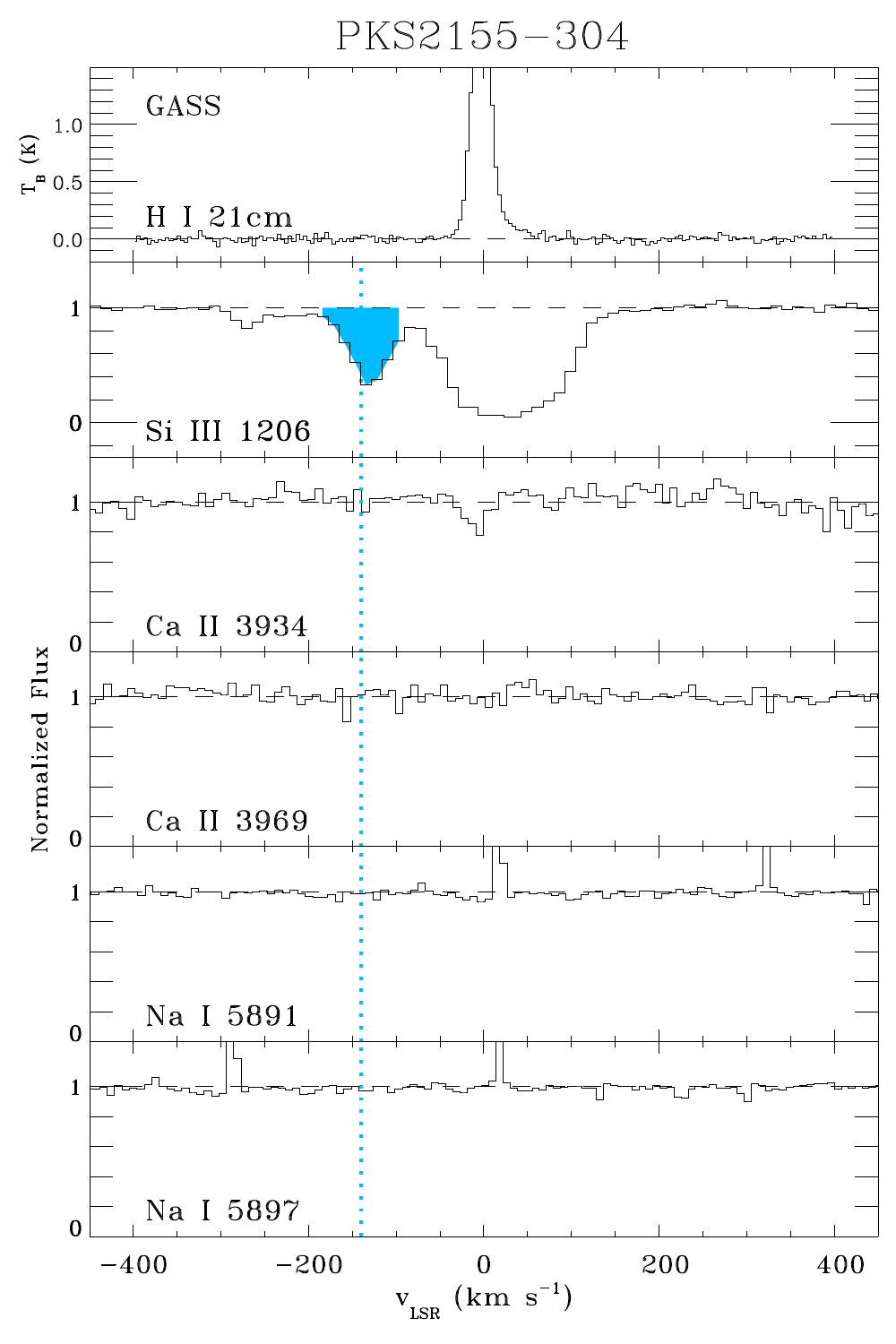}{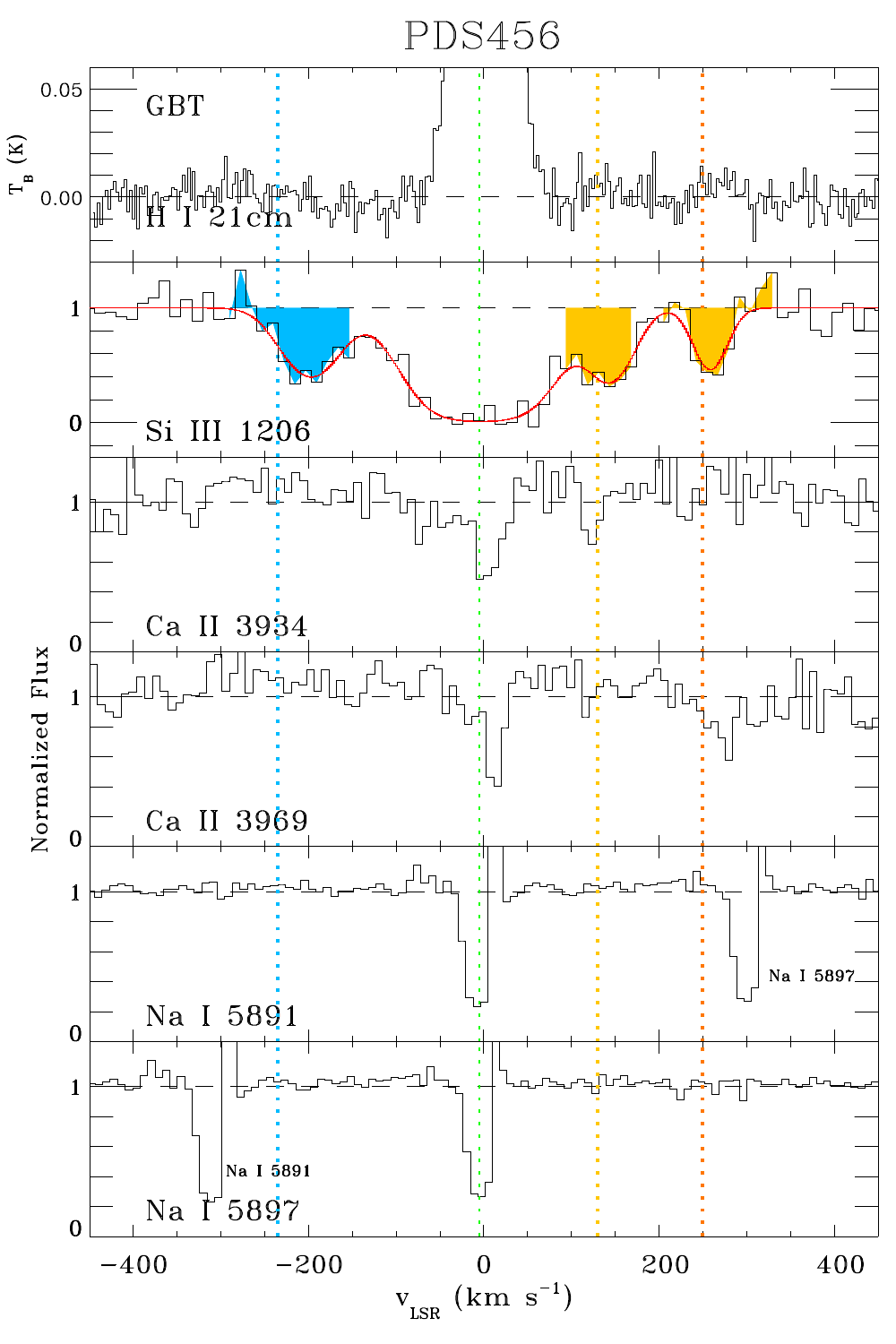}
    \caption{SALT/HRS profiles of \ion{Ca}{2} and \ion{Na}{1} absorption shown alongside the \hi\ 21\,cm GASS profiles and the \ion{Si}{3} profiles from \hst/COS, for two GC  directions, \object{PKS2155-304} and \object{PDS456}. 
    The non-detection of \ion{Ca}{2} and \ion{Na}{1} absorption in the HVCs provides upper limits on the column density of these species. Note that the \ion{Na}{1} profiles toward \object{PKS2155-304} show airglow emission near 0~\kms, but this does not affect the signal at the HVC velocity.}
    \label{fig:salt}
\end{figure*}

We obtained high-resolution optical spectroscopy of two of the AGN in our full GC  sample,
\object{PKS2155-304} and \object{PDS456}, 
with the High-Resolution Spectrograph \citep[HRS;][]{Crause2014} on the Southern African Large Telescope (SALT). These two targets were chosen for optical observations because of their favorable declination for SALT observations and their optical magnitudes. These data were taken under SALT observing program 2016-1-SCI-006 using the HRS high-resolution mode (4 \kms\ FWHM), and were reduced using the standard HRS pipeline \citep{Crawford2015}. The data have been rebinned by five and eight pixels, respectively. The optical bandpass contains two doublets of interest for studying the GC outflow: the \ion{Ca}{2} $\lambda \lambda$ 3933, 3965 doublet (the H \& K lines) and the \ion{Na}{1} $\lambda\lambda$5891, 5897 doublet (the D1 \& D2 lines). Both the \ion{Ca}{2} and \ion{Na}{1} are trace ionization states of cool neutral gas, and both require a high column density of \ion{H}{1} to be detectable. The optical lines are not detected in the GC HVCs in these two directions, providing upper limits on the column density of \ion{Ca}{2} and \ion{Na}{1}. The optical absorption profiles are shown in Figure~\ref{fig:salt}. The \ion{Na}{1} lines are contaminated by airglow, but the \ion{Ca}{2} lines provide useful constraints on the absorbing column in the HVCs under study.

\end{document}